**Rapidly evolving in humans topologically associating domains**

Gennadi V. Glinsky[1, 2]

[1] Institute of Engineering in Medicine

University of California, San Diego

9500 Gilman Dr. MC 0435

La Jolla, CA 92093-0435, USA

Correspondence: gglinskii@ucsd.edu

Web: http://iem.ucsd.edu/people/profiles/guennadi-v-glinskii.html

[2]The Stanford University School of Medicine,

Medical School Lab Surge Bldg., Room P214,

1201 Welch Road, Stanford, CA 94305-5494, USA

Correspondence: gglinsky@stanfrod.edu

**Running title:** Evolution of human-specific regulatory networks

**List of abbreviations**

5hmC, 5-Hydromethylcytosine

CTCF, CCCTC-binding factor

DHS, DNase hypersensitivity sites

FHSRR, fixed human-specific regulatory regions

GRNs, genomic regulatory networks

HAR, human accelerated regions

hCONDEL, human-specific conserved deletions

hESC, human embryonic stem cells

HSGRL, human-specific genomic regulatory loci

HSNBS, human-specific NANOG-binding sites

HSTFBS, human-specific transcription factor-binding sites

LAD, lamina-associated domain

LINE, long interspersed nuclear element

lncRNA, long non-coding RNA

LTR, long terminal repeat

MADE, methylation-associated DNA editing

mC, methylcytosine

mESC, mouse embryonic stem cells

NANOG, Nanog homeobox

nt, nucleotide

POU5F1, POU class 5 homeobox 1

PSDS, partial strand displacement state

TAD, topologically associating domains

TE, transposable elements

TF, transcription factor

TSC, triple-stranded complex

TSS, transcription start sites

SE, super-enhancers

SED, super-enhancer domains

sncRNA, small non coding RNA

**Abstract**

Thousands of candidate human-specific genomic regulatory loci (HSGRL) have been identified, supporting the idea that unique to human phenotypes result from human-specific changes to genomic regulatory networks (GRNs). A notable common feature of HSGRL is a predominant location within non-protein coding sequences. A significant void is the lack of a genome-wide view on diverse families of HSGRL within the context of the principal regulatory structures of the interphase chromatin, namely topologically-associating domains (TADs) and specific sub-TAD structures termed super-enhancer domains (SEDs). Genome-wide proximity placement analysis of 10,598 HSGRL revealed that 0.8%-10.3% of TADs contain more than half of HSGRL. Of the 3,127 TADs in the hESC genome, 24 (0.8%); 53 (1.7%); 259 (8.3%); and 322 (10.3%) harbor 1,110 (52.4%); 1,936 (50.9%); 1,151 (59.6%); and 1,601 (58.3%) HSGRL sequences from four distinct families, respectively. TADs that are enriched for HSGRL and termed rapidly-evolving in humans TADs (revTADs) manifest distinct correlation patterns between HSGRL placements and recombination rates. There are significant enrichment within revTAD boundaries of hESC-enhancers, primate-specific CTCF-binding sites, human-specific RNAPII-binding sites, hCONDELs, and H3K4me3 peaks with human-specific enrichment at TSS in prefrontal cortex neurons ($p < 0.0001$ in all instances). In hESC genome, 331 of 504 (66%) of SE-harboring TADs contain HSGRL and 68% of SEs co-localize with HSGRL, suggesting that HSGRL rewired SE-driven GRNs within revTADs by inserting novel and/or erasing existing regulatory sequences. Consequently, markedly distinct features of chromatin structures evolved in hESC compared to mouse: the SE quantity is 3-fold higher and the median SE size is significantly larger; concomitantly, the TAD number is increased by 42% while the median TAD size is decreased ($p=9.11E-37$). Present analyses revealed a global role for HSGRL in increasing both quantity and size of SEs and increasing the number and size reduction of TADs, which may facilitate a convergence of TAD and SED architectures of interphase chromatin and define a trend of increasing regulatory complexity during evolution of GRNs.

**Introduction**

The genetic and molecular basis defining unique to human phenotypic features remains largely unknown. The near-comprehensive catalogue of changes within protein-coding genes that occurred during human evolution suggested that these changes alone cannot explain all unique to human traits (Chimpanzee Sequencing and Analysis Consortium, 2005; Green et al., 2010; Meyer et al., 2012; Prüfer et al., 2012; 2014; Fu et al., 2014), thus supporting the hypothesis that unique to human phenotypes result from human-specific changes to genomic regulatory sequences (King and Wilson, 1975).

Extensive search for human-specific genomic regulatory loci (HSGRL) revealed thousands candidate HSGRL, a vast majority of which is residing within non-protein coding genomic regions (McLean et al., 2011; Shulha et al., 2012; Konopka et al., 2012; Capra et al., 2013; Marnetto et al., 2014; Glinsky, 2015). Candidate HSGRL comprise multiple distinct families of genomic regulatory elements, including regions of human-specific loss of conserved regulatory DNA termed hCONDEL (McLean et al., 2011); human-specific epigenetic regulatory marks consisting of H3K4me3 histone methylation signatures at transcription start sites in prefrontal neurons (Shulha et al., 2012); human-specific transcriptional genetic networks in the frontal lobe (Konopka et al., 2012); conserved in humans novel regulatory DNA sequences designated human accelerated regions, HARs (Capra et al., 2013); fixed human-specific regulatory regions, FHSRR (Marnetto et al., 2014); and human-specific transcription factor-binding sites, HSTFBS (Glinsky, 2015). Experimentally testable hypotheses on biogenesis and initial evidence of putative human-specific biological functions for several HSGRL families have been reported.

Recent genome-wide computational meta-analysis of TFBS in human embryonic stem cells (hESC) demonstrated that transposable element-derived sequences, most notably LTR7/HERVH, LTR5_Hs, and L1HS, harbor thousands of candidate HSTFBS (Glinsky, 2015). A majority of these candidate HSGRL appears to function as TFBS for NANOG, POU5F1 (OCT4), and CTCF proteins exclusively in hESC, suggesting their critical regulatory role during the early-stage embryogenesis (Glinsky, 2015). Bioinformatics and proximity placement analyses revealed that hESC-specific NANOG-binding sites are enriched near the protein-coding genes regulating brain size, pluripotency long non-coding RNAs, hESC enhancers, and 5-hydroxymethylcytosine-harboring sequences immediately adjacent to TFBS. Candidate human-specific TFBS are placed near the coding genes associated with physiological development and functions of nervous and cardiovascular systems, embryonic development, behavior, as well as development of a diverse spectrum of pathological conditions such as cancer, diseases of cardiovascular and reproductive systems, metabolic diseases, multiple neurological and psychological disorders (Glinsky, 2015).

Calculating as a baseline the evolutionary expected rate of base substitutions based on the experimentally determined level of conservation between multiple species at the given locus, the statistical significance of differences between the observed substitution rates on a lineage of interest in relation to the evolutionary expected rate of substitutions can be estimated. This method appears particularly effective for identifying highly conserved sequences within noncoding genomic regions that have experienced a marked increase of substitution rates on a particular lineage. It has been successfully applied to humans (Pollard et al. 2006; Prabhakar et al. 2006; Bird et al. 2007), where the rapidly-evolving sequences that are highly conserved across mammals and have acquired many sequence changes in humans since divergence from chimpanzees were designated as human accelerated regions (HAR). Experimental analyses of HARs bioactivity revealed that some HARs function as non-coding RNA genes expressed during the neocortex development (Pollard et al. 2006) and human-specific developmental enhancers (Prabhakar et al. 2008). Consistent with the hypothesis that HARs function in human cells as regulatory sequences, most recent computational analyses and transgenic mouse experiments demonstrated that many HARs represent developmental enhancers (Capra et al., 2013).

A notable void is the absence of a genome-wide integration of the rapidly emerging information on diverse families of candidate HSGRL within the context of the interphase chromosome structures defined by the breakthrough studies of interphase chromatin interactions and chromosome folding patterns in human and mouse cells. Recent pioneering work on the interphase chromosome structures revealed specific, reproducible folding patterns of the chromosome fibers into spatially-segregated domain-like segments. In the mammalian nucleus, beads on a string linear strands of interphase chromatin fibers are folded into continuous megabase-sized topologically associating domains (TADs) that are readily detectable by the high-throughput analysis of interactions of chemically cross-linked chromatin (Dixon et al., 2012; Hou et al., 2012; Nora et al., 2012; Sexton et al., 2012). It has been hypothesized that TADs represent spatially-segregated neighborhoods of high local frequency of intrachromosomal contacts reflecting individual physical interactions between long-range enhancers and promoters of target genes in live cells (Dixon et al., 2012; Gorkin et al., 2014). Definition of TADs implies that neighboring TADs are separated by the sharp boundaries, across which the intrachromosomal contacts are relatively infrequent (Dixon et al., 2012; Gorkin et al., 2014).

The identity of ESC critically depends on continuing actions of key master transcription factors (Young, 2011; Ng and Surani, 2011), which govern the maintenance of the ESC's pluripotency state by forming constitutively active super-enhancers (Hnisz et al., 2013; Whyte et al., 2013; Dowen et al., 2014). The most recent experiments show that regulation of pluripotency in ESC occurs within selected TADs by establishing specific sub-TAD structures, which are designed to isolate super-enhancers and target genes within insulated genomic neighborhoods and designated super-enhancer domains (Dowen et al., 2014). Super-enhancer domains (SEDs) are formed by the looping interactions between two CTCF-binding sites co-occupied by cohesin. In ESC, TADs and SEDs are designed to isolate super-enhancers (SEs) and target genes within insulated genomic neighborhoods to facilitate the precision of regulatory interactions between key genetic elements: SE, SE-driven cell identity genes and repressed genes encoding lineage-specifying developmental regulators (Dowen et al., 2014).

In this contribution, genome-wide proximity placement analyses integrating data on 10,598 DNA sequences of individual regulatory elements comprising four distinct families of candidate HSGRL were performed within the context of the principal regulatory components of the interphase chromosome domain structures of the hESC genome. Results of the present analyses revealed mechanistic insights into structural-functional features of HSGRL that may have contributed to the marked changes of the interphase chromatin structure, which are exemplified by the increased number and reduced size of TADs. Concomitant HSGRL-associated increase of both the number and size of SEs points to the increasing regulatory complexity due to convergence of TAD and SED architectures as one of the main directions of the interphase chromatin structural changes during the evolution of genomic regulatory networks.

**Results**

**Identification of topologically-associating domains rapidly-evolving in the hESC genome**

Using genomic coordinates of 3,127 topologically-associating domains (TADs) in hESC (Dixon et al., 2012), a proximity placement analysis of 10,598 DNA sequences representing four distinct families of candidate human-specific genomic regulatory loci (HSGRL) was performed (**Table 1**). The primary criterion for selection of this set of regulatory DNA sequences was the fact that they were identified in human cells lines and primary human tissues whose karyotype were defined as "normal". Based on the origin and definition of corresponding HSGRL, the four HSGRL families were assigned the following designations:

1) Human accelerated regions (HARs; Capra et al., 2013); 2) Human-specific transcription factor-binding sites (HSTFBS; Glinsky, 2015); 3) hESC-derived fixed human-specific regulatory regions (hESC-FHSRR; Marnetto et al., 2014); 4) DNase hypersensitive sites-derived fixed human-specific regulatory regions (DHS-FHSRR; Marnetto et al., 2014). The number of HSGRL placed within a given TAD was computed for every TAD in the hESC genome and the HSGRL placement enrichment was calculated as the ratio of observed values to expected values estimated from a random distribution model at the various cut-off thresholds (**Table 1**). Regardless of the chosen cut-off thresholds, placement of most HSGRL appears markedly restricted to the small fraction of TADs in the hESC genome (**Table 1**). Notably, a majority of individual sequences of each HSGRL family is placed within 0.8-10.3% of TADs in the human genome (**Table 1**). Of the 3,127 TADs in the hESC genome, 24 (0.8%); 53 (1.7%); 259 (8.3%); and 322 (10.3%) TADs are targeted by 1,110 (52.4%); 1,936 (50.9%); 1,151 (59.6%); and 1,601 (58.3%) individual sequences assigned to DHS-FHSRR, HSTFBS, hESC-FHSRR, and HAR families of HSGRL, respectively (**Table 1**). The genome-wide enrichment factors varied for different HSGRL families ranging from 6 to 16-fold for HARs; 7 to 17-fold for hESC-FHSRR; 30 to 45-fold for HSTFBS; and 43 to 88-fold for DHS-FHSRR ($p < 0.0001$ in all instances; **Table 1**). Based on these observations, TADs manifesting a statistically significant accumulation of HSGRL compared to the random distribution model were defined as the rapidly-evolving in humans TADs (revTADs).

**Follow-up analyses of the sixty revTADs enriched for placement of HARs and HSTFBS**

Subsequent analyses were focused on the revTAD set harboring at least 10 individual regulatory DNA sequences assigned to either or both of two distinct HSGRL families: 2,745 human accelerated regions (HARs) and 3,803 human-specific transcription factor-binding sites (HSTFBS). The emergence of these two HSGRL families most likely is a result of different processes, because a vast majority of HSTFBS (99%) represented by human-specific sequences of regulatory DNA which are located within transposable elements (TE) - derived DNA segments (Glinsky, 2015), whereas HARs represent evolutionary highly conserved sequences that have experienced a marked increase of base substitution rates on a human lineage (Capra et al, 2015). A threshold of ten HSRGL per TAD was chosen for the revTAD selection based on a consideration that it would exceed ~10-fold the expected placement number of individual HSGRL per TAD based on a random distribution model estimates.

In the hESC genome, there are sixty TADs (1.9%) meeting these criteria (**Table 2**), 60% of which (36 revTAD) harbor both HARs and HSTFBS (**Supplemental Table 1**). Notably, fifty of sixty revTADs (83%) assigned to this revTAD set harbor at least one HAR (**Supplemental Table 1**). Fourteen revTADs contain within their boundaries at least ten HARs and no HSTFBS, while ten revTADs harbor at least twelve HSTFBS and no HARs (**Supplemental Table 1**). Placement of both HARs and HSTFBS is markedly enriched in this set of revTADs, significantly exceeding the expected numbers for HARs (7.4-fold; $p < 0.0001$) and HSTFBS (18.8-fold; $p < 0.0001$). Among HSTFBS, human-specific CTCF-binding sites manifest the most pronounced placement enrichment (28.4-fold; $p < 0.0001$).

Next, the placement enrichment estimates were computed for multiple other genomic regulatory elements that were previously implicated as candidate regulatory loci with putative impact on human-specific phenotypes and were not considered during the revTAD selection process. Remarkably, placement of hESC enhancers, primate-specific CTCF-binding sites, human-specific RNAPII-binding sites, regions of human-specific conserved deletions (hCONDELs), and H3K4me3 peaks with human-specific enrichment at transcription start sites (TSS) in prefrontal cortex neurons appears significantly enriched within the revTAD boundaries ($p < 0.0001$ in all instances; **Table 2**). Placement of H3K27ac peaks with human-specific enrichment in embryonic limb at E33 stage of human embryogenesis (Cotney et al., 2013) is significantly higher in the revTADs than expected by chance alone (**Table 2**). However, no increase of placements was observed for H3K27ac peaks with human-specific enrichment in embryonic limb at the later stages of embryogenesis, including E37; E41; and E44 stages (data not shown). These results seem to implicate the enhancers and promoters that are engaged during the first five weeks of human embryogenesis in limb development as putative targets for human-specific regulatory elements residing within the revTADs.

One notable exception was the lack of significant placement enrichment for genes comprising human-specific transcriptional genetic networks in the frontal lobe (**Table 2**), which were defined based on the analyses of adult brain tissues (Konopka et al., 2012). However, the *FOXP2* gene encoding one of the principal transcription factors presumably contributing to the human-specific transcriptional control of these networks (Konopka et al., 2012) and previously implicated in evolution of human language and cognition, is residing within the revTAD harboring 12 HAR sequences, one human-specific NANOG-binding site, and 22 primate-specific TFBS, including ten primate-specific CTCF-binding sites. Interestingly, the promoter of the *FOXP1* gene, which can form heterodimers with FOXP2 to regulate transcription and has been implicated in language impairment, intellectual disability, and autism, is also located within another revTAD harboring 10 HAR sequences and 17 primate-specific TFBS, including eight primate-specific CTCF-binding sites. One of the well-known FOXP2 target genes, *LMO4*, is also located within yet another revTAD harboring 10 HAR sequences and 17 primate-specific TFBS, including seven primate-specific CTCF-binding sites. Several lines of experimental evidence strongly argue that LMO4 plays an important role in regulation of asymmetrically developed cognitive processes in humans such as language (Konopka et al., 2012). Nevertheless, the proximity placement analysis does not support the hypothesis that a majority of genes comprising human-specific transcriptional networks in adult brain are located within revTAD regions of human genome. This conclusion is consistent with the previous observations that HSGRL are placed in close proximity to genes having important regulatory functions during the early embryogenesis (Glinsky, 2015).

**Correlation screens revealed distinct patterns of associations between individual members of HSGRL families residing within the revTADs**

At the next stage of the revTAD analysis, a series of correlation screens was performed to determine the relationships between the individual HSGRL residing within the revTAD boundaries (**Figure 1**). To this end, the numbers of individual members of each HSGRL family and primate-specific TFBS located within the boundaries of each revTAD were calculated and corresponding correlation coefficients were computed. Notably, the placement patterns of HSTFBS and individual members of HSTFBS family manifested highly significant positive correlations with the number of primate-specific CTCF-binding sites located within the revTAD boundaries (**Figure 1B**). The most significant positive correlation coefficients were observed for human-specific TFBS and the weakest correlation was recorded for HSTFBS and non-human primate-specific CTCF-binding sites.

In striking contrast, the significant inverse correlations were documented between the placement patterns of HARs and primate-specific CTCF-binding sites residing within the revTADs (**Figure 1C**). The most significant negative correlation coefficients were observed between the placement numbers of HARs and human-specific TFBS and the weakest inverse correlation was recorded between HARs and non-human primate-specific CTCF-binding sites.

The results of these analyses suggest that placement and/or retention patterns of HARs and HSTFBS within the revTADs are guided and governed by distinct mechanisms. Placement and/or retention of HSTFBS appear to follow the CTCF-binding sites' patterns, whereas locations of HARs seem to favor the revTAD regions harboring relatively fewer CTCF-binding sites resulting in highly significant inverse correlation between placement patterns of HARs and HSTFBS within the revTADs (**Figure 1D**).

**Distinct correlation profiles of HSGRL and recombination rates within the revTADs distinguish placement patterns of HARs and HSTFBS**

It has been reported that a prevalent mode of mutations in HARs is base substitutions that change weak (A, T) into strong (G, C) base pairs, which may occur during meiotic recombination as a result of a biochemical bias towards strong G/C alleles during the mismatch repair of heteroduplex DNA molecules (Kostka et al., 2012). Consistent with this notion, the enrichment of GC-biased substitutions of DNA sequences near recombination hotspots and a significant correlation between GC bias and recombination rate in the human genome have been reported (Katzman et al., 2011). Direct measurements of fine-scale recombination rates in genomic regions surrounding hominid accelerated conserved regions demonstrated significantly higher mean recombination rate estimates for 30 Kb DNA segments around HARs (Freudenberg et al., 2007). These observations prompted detailed examination of recombination rates within the revTADs. Recombination rates were downloaded from the HapMap Project (The International Hapmap Consortium, 2007) and the number of DNA segments with the recombination rates of 10 cM/Mb or greater were identified for each revTAD. The results were plotted for visualization of spatial distributions (**Figures 2A, B**) and corresponding correlation coefficients were computed (**Figures 2C, D**). In these analyses, the total numbers of recombination hotspots having recombination rates of 10 cM/Mb or greater within boundaries of a given revTAD were determined and designated as the recombination scores (see Materials and Methods for details).

Significant inverse correlations were observed between recombination scores and the numbers of HSTFBS residing within the revTADs (**Figure 2C**), whereas no significant correlation was recorded between recombination scores and non-human primate-specific CTCF-binding sites. In striking contrast with HSTFBS, a highly significant positive correlation was observed between recombination scores and the numbers HARs located within the revTADs (**Figure 2D**). Interactions of DNA strands are required for recombination process. Consistent with this notion, a significant positive correlation was recorded between the numbers of intrachromosomal contacts observed within a given revTAD region and the corresponding recombination scores (**Figure 2D**). This observation offers an opportunity to analyze the relationships between recombination rates and placement of distinct HSGRL in sub-groups of revTADs segregated based on the mean values of cumulative numbers of intrachromosomal contacts. Intrachromosomal contacts represent analytically and technically distinct set of variables, which has been measured genome-wide in entirely independent set of experiments by design, execution, and technical protocols (Jin et al., 2013). The strikingly distinct association profiles between recombination rates and placement patterns of either HARs or HSTFBS were documented when the mean values of corresponding variables were computed and analyzed for the revTAD sets, which were segregated into quartiles based on the cumulative values of numbers of intrachromosomal contacts observed within the regions (**Figure 2D**). The median recombination scores for the revTADs placed in the top quartile based on the quantity of observed intrachromosomal interactions was 40-fold greater compared to the median recombination score of the revTADs assigned to the bottom quartile (**Figure 2D**).

The results of these analyses are consistent with the hypothesis that placements of HARs and HSTFBS within the revTADs are associated with distinct molecular processes and support the idea connecting the biogenesis of HARs with high recombination rates. It has been demonstrated that HSTFBS are located almost exclusively within TE – derived DNA sequences (Glinsky, 2015), strongly implicating activity of TE in biogenesis of HSTFBS. Present observations of significant inverse correlations between the HSTFBS placement numbers and recombination rates within the revTADs are highly congruent with this hypothesis, because TE insertions are known to evade the genomic regions with high recombination rates (Rizzon et al., 2002).

**Markedly distinct features of super-enhancers in the hESC genome compared with mESC**

The sustained expression of key cell identity genes and repression of genes encoding lineage-specifying developmental regulators is essential for maintaining ESC identity and pluripotency state. These processes are governed by the master TFs OCT4 (POU5F1), SOX2, and NANOG (OSN), that function by establishing SEs regulating cell identity genes, including master TFs themselves ((Hnisz et al., 2013; Whyte et al., 2013). The spatial relationships between SEs and HSGRL in the hESC genome within the context of TADs have not been investigated. To this end, the placement enrichment analysis was carried out to identify all TADs in hESC genome that harbor SEs and examine the association of SEs and HSGRL. There are 504 TADs (16%) harboring 642 SEs (94%) in the hESC genome. Remarkably, significant placement co-enrichments were observed between SEs, HSTFBS, and HARs residing within the boundaries of 279 TADs in the hESC genome (**Table 3**). In total, 279 TADs (8.9%) harbor 369 SEs (57.5%) that are co-localized with 300 HSTFBS and 564 HARs. When other HSGRL families listed in the **Table 1** were considered in co-localization analyses, 331 TADs (10.6%) harboring 436 SEs (67.9%) were found to co-localize with HSGRL.

In human genome, there are approximately 7,000 high-confidence hESC-enriched enhancers (Xie et al., 2013; Hnisz et al., 2013), less than 10% of which are defined as SEs (Hnisz et al., 2013). Placement of high-confidence hESC enhancers is significantly enriched within the revTADs (**Table 2**), implying possible genome-wide associations of HSGRL and hESC-enriched enhancers. Indeed, such associations became apparent when genomic co-localization analyses were performed to asses numbers of HSGRL and hESC-enriched enhancers residing within the boundaries of TADs. These analyses revealed that there is a significant direct correlation between the numbers of hESC-enriched enhancers and HARs that are located within the same revTADs (**Figure 3**). Genome-wide placement and/or retention of HARs appears enriched within hESC enhancer-harboring TADs and seems to favor TADs containing larger numbers of hESC-enriched enhancers (**Figure 3**).

Co-localization of a majority of SEs with HSGRL suggest that HSGRL may affect the structural-functional features of SEs in the hESC genome. Consistent with this hypothesis, genomes of human and mouse ESC manifest markedly distinct features associated with SE structures and functions. Despite strikingly similar genome sizes and numbers of protein-coding genes, hESC genome contains 3-fold more SEs compared to mouse: there are 684 SEs in the genome of hESC (Hnisz et al., 2013) and 231 SEs in the genome of mESC (Whyte et al., 2013). In the genome of hESC, only twenty-five of mESC SEs (11%) are represented as conserved orthologous sequences having genomic architecture of SEs, suggesting that 89% of mESC SEs lost the SE features in humans. Furthermore, 96% of SEs in the genome of hESC acquired structural-functional features of SEs during evolution after Euarchonta and Glires split 88 million years ago. The median size of SEs in hESC appears significantly larger compared to the median SE size in mESC ((9,589 bp versus 8,667 bp, respectively; $p = 0.017$). Detailed size distribution analyses demonstrated that accumulation of SEs in hESC genome is associated with increased number of large SEs and decreased number of small SEs compared to mouse genome: there are 7-fold increase of very large SEs having size more than 30 Kb, consistent ~3-fold increase of SEs having size range from 2 – 30 Kb, and a marked 38-fold depletion of small SEs having size less than 2 Kb (**Table 4; Figure 4**).

Collectively, these data indicate that structural-functional features of SEs are markedly distinct in the hESC genome compared to mouse, which appear associated with the enrichment of HSGRL within SE-harboring TADs. It will be of interest to determine whether the HSGRL placement within SE-harboring TADs exerts biologically-meaningful effects on SE functions. Targeted placements and/or retention of HSGRL may increase density of enhancer elements within selected TADs, which would result in the merger of conventional enhancer units into super-enhancer structures containing the exceptionally high level of transcription-enabling apparatus to drive and continually maintain high expression of associated target genes. This idea is supported by the findings that nearly 60% (1571) of the HARs overlap at least one of the common markers of enhancer activity in human cells (Capra et al., 2013).

**TAD structural features are markedly altered in the hESC genome compared with mESC**

There are 3,127 TADs in the hESC genome, which is 42% more than 2,200 TADs in the genome of mESC (Dixon et al., 2012). Coincidently, there are 87,883 CTCF-binding sites in hESC (Kunarso et al., 2010), 29,018 (33%) of which represent primate-specific CTCF-binding sequences (Glinsky, 2015). The median size of TADs in hESC is significantly smaller compared to the median size of TADs in mESC (680 Kb versus 880 Kb; $p = 9.11E-37$). Detailed size distribution analyses of TADs in human and mouse ESC (**Figure 4; Table 5**) revealed that there is a ~ 2-fold depletion of large TADs having size > 2,000 Kb and consistently increased numbers of medium-size and small TADs having size range from 100 - 1,000 Kb (**Table 5**). Structural-functional features of the revTADs in the hESC genome appear markedly distinct from the TADs of the orthologous sequences in the mouse ESC genome. Of the 60 revTAD listed in the **Table 2**, nineteen (32%) are placed within the primate-specific sequences that failed to align to the mouse reference genome database sequence. Remaining revTADs appear to evolve by erasing the existing and establishing new domain boundaries within orthologous DNA sequences via two distinct mechanisms: 1) domain & boundary crossing (hESC TADs appear to cross boundaries of 2 to 4 orthologous TADs in mESC genome); and 2) domain shrinking & boundary creation (smaller hESC TADs are placed within the boundaries of larger orthologous TAD sequences in mESC genome).

The above considerations prompted additional analyses of relationships between the SEs and TADs in the genomes of human and mouse ESC (**Figure 5**). These analyses revealed that in hESC there are significant direct correlations between the size of TADs and SE's span defined as a number of bp between the two most distant SEs located within a given TAD (**Figure 5A**). Similar trends were observed in the mESC genome, however, the correlation coefficient values were not statistically significant. Correlation patterns observed for the 60 revTADs for placements of hESC-enriched enhancers, HARs, HSTFBS, and size of TADs were validated on a larger set of 147 revTADs (**Figure 5B**). Genome-wide, the highly significant direct correlation was discovered between the size of TADs and the number of hESC-enriched enhancers located within TADs (**Figure 5C**). These observations are conceptually coherent because TAD boundaries were inferred from the relative prevalence and directionality of interchromosomal interactions along the chromosome length (Dixon et al., 2012), which are predominantly mediated by the enhancers' activities and detected as enhancer-promoter and enhancer-enhancer interactions in the Hi-C analyses. It will be of interest to determine experimentally whether the size and boundaries of the adjacent and neighboring TADs can be altered by increasing the density, structure, and activity of the resident enhancers.

**Potential mechanisms of HSGRL-mediated effects on principal regulatory structures of interphase chromatin**

Present analyses provide a conceptual framework for understanding genome-scale regulatory changes during evolution within the context of the principal regulatory structures of the interphase chromatin to reflect an apparent trend toward increasing complexity of genomic regulatory networks (**Figure 6**). Experimental observations at the foundation of building blocks of the genome's evolution model (GEM) are presented in the previous sections and additional considerations are focused on the analyses of potential contributions of SEs and hESC-enriched enhancers to these processes.

According to GEM, one of the key elements of the evolution of genomic regulatory networks is the creation of new enhancer elements (**Figure 6**). Conventional enhancers comprise discrete DNA segments occupying a few hundred base pairs of the linear DNA sequence and harboring multiple TFBS. SEs consist of clusters of conventional enhancers that are densely occupied by the master transcription factors and Mediator (Whyte et al., 2013). Therefore, it is logical to expect that creation of new TFBS and increasing density of TFBS would increase the probability of the emergence of new enhancer elements. Creation of new enhancers and increasing their density would facilitate the emergence of new SE structures. This sequence of events would imply that evolutionary time periods required for creation of TFBS, enhancers, and SEs are shortest for TFBS, intermediate for enhancers, and longest for SEs. To test this assumption, estimates of creation time periods for enhancers and SEs in the hESC genome were calculated (**Table 6**) and compared to the estimates of creation rates of TFBS (Glinsky, 2015). Consistent with the model expectations, the estimated creation time is markedly longer for SEs compared to enhancers for conserved (8-fold), primate-specific (17-fold), and human-specific (63-fold) regulatory sequences (**Table 6**). Furthermore, the estimated creation time periods for enhancers appear several fold longer compared to the creation time estimates for TFBS sequences in both chimpanzee and humans (Glinsky, 2015). One of the intriguing results of this analysis is that 87% of new SEs and 75% of new enhancers in the hESC genome were created within conserved sequences (**Table 6**). Similarly, a majority (67%) of new binding sites for NANOG (59,221 of 88,351 TFBS) and CTCF (58,865 of 87,883 TFBS) proteins in the hESC genome compared with mESC are located within conserved sequences (**Supplemental Figure S3**).

Notably, the estimated creation time appears accelerated in humans compared to chimpanzee for both SEs (7-fold) and enhancers (27-fold), suggesting that human genomes were acquiring new regulatory elements and increasing the regulatory complexity and precision of genomic regulatory networks at the markedly accelerated pace.

CTCF-binding sites play a crucial role in defining the TAD boundaries (Dixon et al., 2012; Li et al., 2013) and in establishing the SED architecture (Dowen et al., 2014). Knocking down the expression of CTCF and cohesin genes to reduce their protein concentrations results in a decrease in the number of intra-TAD chromatin loops (Sofueva et al., 2013; Seitan et al., 2013; Zuin et al., 2014), increase in the number of inter-TAD chromatin interactions crossing the TAD boundaries (Zuin et al., 2014), and chromatin compaction (Tark-Dame et al., 2014). Several lines of evidence are in agreement with the hypothesis that creation of new CTCF-binding sites may have contributed to the rewiring of genomic regulatory networks during primate evolution:

i) In the hESC genome, 29,018 of 87,883 (33%) CTCF-binding sites represent primate-specific sequences (Kunarso et al., 2010; Glinsky, 2015);

ii) Of the 23,709 constitutive CTCF-binding sites implicated in defining TAD boundaries in the human genome (Li et al., 2013), 6,787 sites (28.6%) failed alignment to the mouse genome, suggesting that they represent primate-specific sequences;

iii) Two hundred eighty seven HARs (10.5%) are located within 10 Kb of the 336 constitutive CTCF-binding sites;

iv) Five hundred seventy one HARs (20.8%) are located within 1 Kb of 953 CTCF-binding sites;

v) Activation of retrotransposons and lineage-specific repeat-driven dispersion of CTCF-binding sites has produced species-specific expansions of CTCF binding in mammalian genomes and these new CTCF-binding sites function as chromatin domain insulators and transcriptional regulators (Schmidt et al., 2012).

Detailed proximity placement and HAR/TFBS co-localization analyses identified 123 HARs located within 1 Kb from 127 high-confidence CTCF-binding sites, which were validated in 23 different human cells lines. Notably, 123 of 127 (97%) high-confidence CTCF-binding sites represent overlapping CTCF/RAD21-binding sites, which is consistent with their putative regulatory role in establishing TAD boundaries and/or SED architecture. A significant majority of 123 HARs located near high-confidence overlapping CTCF/RAD21-binding sites harbor at least one TFBS in human cells (76 HARs; 62%).

Significant increase of TAD and SED numbers in human genomes compared with mouse indicates that chromatin folding profiles are dramatically altered, reflecting more frequent intrachromosomal segmentation of linear DNA fibers and enhanced likelihoods of formation of isolated intra-segmental looping structures. These changes of chromatin folding profiles and DNA fibers' packaging patterns would impose more stringent requirement on DNA elasticity, particularly, near the TAD and SED boundaries due to bending of DNA double helix. Consequently, changes of DNA sequence and/or chromatin structure may be necessary to accommodate these new structural requirements. It has been reported that SINE repeats are enriched at the TAD boundaries (Dixon et al., 2012), suggesting that insertion of repetitive elements may contribute to changes of DNA elasticity near putative bending sites. Consistent with this hypothesis, a survey of SED sequences in hESC revealed a systematic significant enrichment of Alu elements within ~5 Kb windows near SED boundaries (**Figure 7**). Clusters of closely-spaced sequences of at least three Alu elements belonging to most ancient AluJ (~65 million years old), second oldest AluS (~30 million years old), and currently active modern AluY sub-families (Bennett et al., 2008) were observed frequently, suggesting that placement and/or retention of Alu elements at these sites were occurring for millions of years and continues at the present time.

Taken together with the recent report that nearly 60% of the HARs overlap at least one of the common markers of enhancers in human cells (Capra et al., 2013), these observations strongly argue that creation of new enhancer elements is one of the key events defining evolution of genomic regulatory networks. Marked acceleration of the TFBS and enhancers' creation processes in humans may have contributed to genome-scale rearrangements of principal regulatory structures of the interphase chromatin, leading to the increased complexity and enhanced precision of genomic regulatory networks and emergence of human-specific phenotypes. Frequent locations of Alu sequences near the putative DNA bending sites suggest their potential role in regulation of the elasticity of chromatin fibers by reducing nucleosome placements due to the low affinity of Alu elements to nucleosome binding (Huda et al., 2009) and/or facilitating strand invasion reactions between neighboring DNA strands leading to Alu/Alu recombination events (Lee at al., 2015; Morales et al., 2015). This model is in agreement with the observations that Alu elements appear preferentially retained in GC-rich and gene-rich regions of the human genome (Lander et al., 2001).

**Conservation patterns of HSGRL in individual human genomes**

All analyses conducted so far were performed using the reference genome databases and not the individual human genomes. To address this limitation, the assessment of conservation of HSGRL in individual genomes of 3 Neanderthals, 12 Modern Humans, and the 41,000-year old Denisovan genome (Reich et al., 2010; Meyer et al., 2012) was carried-out by direct comparisons of corresponding sequences retrieved from individual genomes and the human genome reference database (http://genome.ucsc.edu/Neandertal/ ). Full-length sequence alignments with no gaps of the individual genome sequences to the corresponding sequences in the human genome reference databases of both hg18 and hg19 releases were accepted as the evidence of sequence conservation in the individual human genome.

Sequences of all analyzed to date HSGRL appear conserved in individual human genomes (**Figure 8** and **Supplemental Figure S4**), albeit a significant inter-individual variability in degree of conservations of specific HSGRL is apparent. Conservation of all HSGRL are consistently at the lowest level in the Neanderthals' genome, suggesting that creation and/or retention rates of HSGRL are enhanced in Modern Humans. The results of these analyses indicate that sequences of HSGRL with assigned biochemical functions, e.g., specific TFBS or Lamin B1 (LMNB1)-binding sites, which are residing within HARs exhibit markedly higher conservation levels compared to sequences of HARs harboring the corresponding HSGRL. This conclusion remains valid for HSGRL sequences with assigned specific biochemical or biological functions that were associated with HARs by proximity placement analyses (**Figure 8A, B**). HSGRL sequences manifesting the relatively high conservation levels in the Neanderthals' genome appear most conserved in the individual genomes of Modern Humans as well, including the 41,000-year old Denisovan genome (**Figure 8B**).

Consistent patterns of significant direct correlations between conservation profiles of distinct seemingly unrelated HSGRL sequences in individual human genomes were observed (**Figure 8C** and **Supplemental Figure S4**). One possible interpretation of these observations is that HSGRL conservation patterns reflect intrinsic features of the individual human genome, integration of which can gauge the overall capacity of a genome to create and retain HSGRL. This idea was tested by calculating for each individual human genome the genomic fitness scores integrating into a single numerical value sequence conservation data of 909 HSGRL (**Figure 8D**). Notably, individual human genomes appear markedly distinct based on the results of this analysis with the lowest genomic fitness score of 0.9 in Neanderthals, followed by the scores of 1.89; 2.22; and 2.27 for individual genomes of Native American, Denisova cave, and Mongolian subjects, respectively. On the other end of the spectrum, the genomic fitness scores of 4.51; 4.72; and 4.73 were obtained for genomes of Yoruba (West Africa), French (Western Europe), and San (Southern Africa) individuals, respectively.

These observations are highly congruent with the recent definition of structural variations-free (fixed) human-specific regulatory regions (FHSRR) based on the stringent formal analysis of the HSGRL intra-species patterns of variations within human population using exome and full genome sequencing database of 1,092 individuals from the 1000 Genomes Project Consortium (Abecasis et al., 2012; Marnetto et al., 2014). Collectively, the consistent evidence of HSGRL conservation in individual human genomes are in accord with the hypothesis of their putative functional role in defining human-specific phenotypes.

**Working models of *NANOG*, *POU5F1*, and *POU3F2* super-enhancers' domains in hESC**

One of the potential practical utilities of the outlined concepts is the opportunity to use the available genomic information for building the experimentally testable working models of SEDs and TADs, which may be suitable for precise molecular and genetic definitions of critical structural elements of human-specific regulatory networks. To test this notion, the GEM principles (**Figure 6**) were utilized as a blueprint to build working models of *NANOG*, *POU5F1*, and *POU3F2* super-enhancers' (SEDs) domains in hESC (**Figure 9**;

**Supplemental Figure S5**) and compare them to the corresponding SED structures recently identified in the mESC genomes (Dowen et al., 2014).

To identify the putative SED boundaries, a systematic search was conducted for nearest overlapping CTCF/cohesin sites located next to borders of the two *NANOG* super-enhancers (Hnisz, et al., 2013). One of the notable features of the human *NANOG* SED is the presence of a large primate-specific region, which places the down-stream *NANOG* SED boundary within lamina-associated domain (LAD) harboring the primate-specific overlapping CTCF/cohesin site (**Figure 9A**). Modeling of the human *NANOG* SED reveals that there are at least three possible designs of SED architecture that would incorporate both the *NANOG* SE and the target gene *NANOG* within the isolated genomic neighborhood of the looping structure formed as a result of interactions between two overlapping CTCF/cohesin sites (**Figure 9A**). Two of SED design variants would include only one *NANOG* SE and would require the interactions between the primate-specific and conserved overlapping CTCF/cohesin sites enabling formation of the isolated genomic neighborhood loops of relatively smaller sizes (**Figure 9A**; **Supplemental Figure S5**). The largest *NANOG* SED looping structure would result from interactions between two primate-specific overlapping CTCF/cohesin sites and it would incorporate both *NANOG* SEs within SED (**Figure 9A**). Notably, the postulated *NANOG* SED sequence harbors five primate-specific NANOG-binding sites previously identified in the hESC genome (Kunarso et al., 2010; Glinsky, 2015), which may indicate that NANOG protein exerts more efficient control over its own locus in primates' genomes. In the mESC genome, the *NANOG* SED is markedly smaller occupying just 52 Kb and harboring only a single *NANOG* SE (Dowen et al., 2013). The entire SED region down-stream of the *NANOG* gene, which contain the *NANOG* SE sequence, has no orthologous sequences in primate and human genomes and appears replaced by the large primate-specific sequence in the human genome (**Figure 9A**; **Supplemental Figure S5**).

Similarly dramatic rearrangements of the SED architecture in the hESC genome are apparent when additional examples of the SED structures insulating SEs regulating *POU5F1* and *POU3F2* target genes were analyzed (**Figure 9B**; **Supplemental Figure S5**). In all instances, the genomic looping structure-enabling interactions between two overlapping CTCF/cohesin sites would require the involvement of at least one primate-specific overlapping CTCF/cohesin-binding site. The *POU5F1* SED region harbors one human-specific and three primate-specific NANOG-binding sites (**Figure 9B**; **Supplemental Figure S5**), whereas the *POU3F2*

SED contains ten hESC-enhancers and twenty-eight primate-specific TFBS for NANOG (15 sites), POU5F1 (6 sites), and CTCF (7 sites) proteins (**Supplemental Figure S5**). The example of *POU3F2* SED is of particular interest because it is located within one of the rapidly-evolving in humans TADs and, according to recent experiments, the *POU3F2* gene plays a pivotal role in the highly efficient conversion of human pluripotent stem cells as well as non-neural human somatic cells into functional neurons (Pang et al., 2011; Wapinski et al., 2013). Of note, sequences near the overlapping CTCF/cohesin sites, interactions between which bring about the SED structures, appear to have a common structural element comprising two conserved in individual human genomes LMNB1-binding sites adjacent to CTCF/cohesin sites (**Figure 9C; Supplemental Figure S5**). These LMNB1-binding sites may serve as anchors to nuclear lamina for DNA sequences harboring the overlapping CTCF/cohesin sites, thus thermodynamically enhancing the likelihood of creation and/or stabilizing the maintenance of the base of SED structures.

Applying the GEM principles (**Figure 6**) as a guidance to infer the putative SED and TAD structures, it is possible to build the working models of the multidimensional chromatin architecture for large genomic regions spanning 2.54-3.26 Mb and containing 3-4 TAD structures (**Figures 10 & 11; Supplemental Figure S6**). Key elements of this design strategy entails the following considerations:

i)  Defining the positions of SED bases by the overlapping CTCF/cohesin sites interacting with each other to create the SED structure;

ii) Mapping the putative bending sites of linear chromatin fibers based on the positions of Alu clusters near the overlapping CTCF/cohesin sites;

iii) Requiring that SEs and their target genes are located within the same insulated chromatin looping structure;

iv) Taking into account the relative strength of the chromatin fiber interactions with nuclear lamina based on placements of corresponding DNA sequences within or outside LADs and/or based on the quantity of LMNB1-binding sites within a region;

v) Strictly adhering to the previously defined genomic positions of TAD and LAD boundaries, ESC enhancers, primate-specific TFBS, and HSGRLs.

Using this strategy, a model has been built representing a global view of multidimensional chromatin folding patterns within the 2.54 Mb genomic region containing the *NANOG*-residing TAD and two adjacent neighboring TADs (**Figure 10).** A more detailed depiction of sequential steps of the model building is presented in the **Supplemental Figure S6**. It seems evident, that one of the principal structural elements defining directionality of the *NANOG* SED looping structure in the human genome is the insertion immediately downstream of the *NANOG* gene of the primate-specific sequence, which is tightly attached to the nuclear lamina as depicted in the **Figure 10** by placement of the corresponding part of the *NANOG* SED within red-colored LAD. The remaining part of the *NANOG*-harboring TAD represents the genomic region located upstream of the *NANOG* SED. It is not attached to the nuclear lamina and contains the *SLC2A3* SED, which is located in its entirety outside of the LAD boundaries (**Figure10**). Further upstream of the *SLC2A3* SED at the boundary region between the *NANOG*-harboring TAD and adjacent upstream TAD, the chromatin fiber re-attaches to nuclear lamina and places within LAD the entire TAD containing 4-5 insulated neighborhood looping structures (**Figure 10**). It will be of interest to determine how transitions of placements of chromatin fibers between positions within and outside LAD would affect the creation of SED structures and performance of insulated within SED genomic regulatory circuitry. One of the near-term practical applications of this approach to multidimensional modeling of SED and TAD structures became apparent when the regulatory elements residing within corresponding TADs and sub-TAD structures were catalogued and reviewed.

**Conserved profiles and uniquely distinct patterns of H3K27ac peaks' distribution within the *NANOG*-associated genomic region in 43 types of human cells and tissues**

One of the characteristic regulatory features systematically associated with hESC enhancers located within four TADs of the *NANOG* region is the presence of multiple binding sites for both EP300 histone acetyltransferase and histone deacetylase 2, HDAC2 (**Figures 10 & 11; Supplemental Figure S6**). Both histone acetyltransferases (HATs) and histone deacetylases (HDACs) have essential and pleiotropic roles in regulation of stem cell self-renewal by maintaining expression of master TFs regulating the pluripotent state and controlling the core transcriptional regulatory networks of embryonic stem cells and their differentiated progenies (Chen et al., 2008; Fazzio et al., 2008; Zhong X, Jin Y, 2009; Li et al., 2012; Jamaladdin et al., 2014). These observations are likely reflect the requirement for constitutive activities of both HATs and HDACs at transcriptionally active regulatory elements, because nucleosome maintenance and turnover are highly dynamic at these loci due to markedly accelerated exchange rates of histones, which appear directly associated with the histone modifications defining the active chromatin regulatory state. For example, H3.3-containing nucleosomes undergo rapid turnover at active enhancers and promoters (Kraushaar et al., 2013). This fast turnover is positively correlated with active chromatin histone modification marks, including H3K4me1, H3K4me3, H3K9ac, and H3K27ac (Kraushaar et al., 2013). In contrast, the nucleosome exchange rate is negatively correlated with repressive chromatin marks H3K27me3 and H3K9me2 at regulatory loci and H3K36me3 modifications within gene bodies (Kraushaar et al., 2013).

It is intriguing that binding sites for HDAC2 but not HDAC1 were detected in this region. Of note, it has been demonstrated that a single allele of *Hdac2* but not *Hdac1* is sufficient to rescue normal mouse brain development in double knockout *Hdac1*$^{-/-}$*Hdac2*$^{-/-}$ mice and that HDAC2 has a unique indispensable role in controlling the fate of neural progenitor cells (Hagelkruys et al., 2014). Forced neuron-specific overexpression of HDAC2, but not HDAC1, reduced dendritic spine density, synapse number, synaptic plasticity, and memory formation in mice (Guan et al., 2009). Consistently, reduction of synapse number and learning impairment of *Hdac2*-overexpressing mice were ameliorated by chronic HDACi treatment (Guan et al., 2009). In the mouse models of neurodegeneration and in brains of patients with Alzheimer's disease, cognitive capacities were severely impaired by the epigenetic blockade of gene transcription at specific genetic loci important for learning and memory, which is mediated by the increased HDAC2 activity at these sites (Gräff et al., 2012). Neurodegeneration-associated memory impairments, abnormal structural and synaptic plasticity, and diminished expression of genes regulating learning and memory were reversed following shRNA-mediated knockdown of HDAC2 overexpression (Gräff et al., 2012). Critical role of HDAC2 in regulation of brain functions is supported by the recent experiments on individual neurons demonstrating that HDAC2 cell autonomously suppresses excitatory activity and enhances inhibitory synaptic function in CA1 pyramidal neurons (Hanson et al., 2013).

Since SE structures in genomes of human cells were defined based on the genomic profiles of H3K27ac peaks (Hnisz et al., 2013), it is reasonable to argue that patterns of SE placements within TAD reflect genomic positions of H3K27ac peaks and should faithfully represent the snap-shot of the balance of continuing HAT and HDAC activities at specific regulatory sites within a region. Based on these considerations, genomic maps of SE distributions were drawn for 43 distinct types of human cells and tissues by retrieving genomic coordinates of SEs from the recently reported comprehensive catalogue of SEs in human body (Hnisz et al., 2013). Regional genomic maps of SE distributions shown in **Figure 11** depict numbers of SEs placed within 3.26 Mb genomic region containing four TADs and three TAD boundaries for individual types of human cells and tissues against the background of SE distribution profiles within the same region in the hESC genome. It is evident that all individual profiles of SE distributions in human cells and tissues, which reflect distribution profiles of H3K27ac peaks, follow the general pattern of SE distributions established in the hESC genome (**Figures 11B & 11C**). One of the notable common elements that emerged in many differentiated cells and tissues is the appearance of SEs within the genomic regions demarcating TAD boundaries in the hESC genome, suggesting that TAD boundaries and TAD structures underwent visible rearrangements in humans' differentiated cells and tissues compared with hESC. Erasing the hESC TAD boundaries during differentiation due to appearance of new cell-type specific SEs is consistent with observed decreasing numbers of TADs in differentiated cells and tissues (**Figure 4**).

Transformation of these data into different types of charts for patterns' visualization (**Figures 11B-E**) revealed that all analyzed human cells and tissues manifest unique profiles of SE/H3K27ac distributions within the 3.26 Mb *NANOG*-associated genomic region. It seems quite remarkable that quantitative distribution profiles of the very small set of markers located within ~0.1% of human genome appear to distinguish 43 distinct types of human cells and tissues. This notion is best illustrated in the **Figure 11E** showing the data set of 43 distinct types of human cells and tissues that were sorted in descending order based on total numbers of SEs located within the 3.26 Mb *NANOG*-associated genomic region. It is clearly visible that most samples are easily distinguishable by the total number of SEs located within the region and samples that have the same numbers of SE's manifest readily distinct SE/H3K27ac distribution profiles (**Figure 11E**). These results seem to point to a stochastic, i.e., random and probabilistic, nature of the underlying causal mechanisms driving the emergence of unique profiles of SE/H3K27ac distributions in differentiated cells and tissues from the singular parental SE/H3K27ac distribution profile in hESC. Consistent with this hypothesis, 45.2% of human cells' and tissues' samples have the higher numbers of SE's placements compared with hESC, whereas the SE's placement numbers are lower compared with hESC in 54.7% of samples (**Figure 11E**). Collectively, this analysis appears to indicate that the emergence of unique cell type-specific patterns of SE/H3K27ac placements may have been triggered by the small bidirectional changes of the balance between the HAT's and HDAC's activities in the genome of hESC that evolved during the differentiation process into arrays of quantitatively distinct patterns of SE/H3K27ac distributions in different types of cells and tissues in human body.

**Putative regulatory consequences of targeted placements of HSGRL affecting expression of protein-coding genes**

Observed increasing regulatory complexity in human genomes associated with targeted placements of HSGRL is consistent with a view that increasing genomic complexity within fixed environments represents a major evolutionary trend in the natural world (Adami et al., 2000). Further analysis suggests that these changes of nuclear regulatory architecture may facilitate the enhanced precision of regulatory interactions between enhancers and target genes within insulated neighborhoods of chromatin regulatory networks. The boundaries of a majority of TADs are shared by the different cell types within an organism, segregating genomes into distinct regulatory units that harbor approximately seven protein-coding genes per TAD (Dixon et al., 2012; Smallwood and Ren, 2013). In contrast, nearly two-third of revTADs (39 of 60 revTADs; 65%) contain five or less protein-coding genes. Among these, eight revTADs harbor just one protein-coding gene and two revTADs contain only non-coding RNA genes. Of the twenty one revTADs harboring more than five protein-coding genes, seven revTADs contain clusters of functionally-related protein-coding genes: 3 revTADS on chr11 harbor 61 genes encoding olfactory receptors; 3 revTADs on chr19 harbor 31 genes encoding zinc finger proteins; and one revTAD on chr20 contains eight beta-defensins' genes. Importantly, the expression of genes residing within the same TAD appears somewhat correlated and some TADs tend to contain predominantly actively transcribed genes while others have mostly repressed genes (Cavalli and Misteli, 2013; Gibcus and Dekker, 2013; Nora et al., 2012). Collectively, these results suggest that observed enrichment of HSGRL placements in specific genomic regions may facilitate the enhanced precision of regulatory functions in human genomes by targeting TADs harboring relatively few protein-coding genes and/or TADs containing clusters of functionally-related protein-coding genes, thus preferentially affecting insulated regulatory networks governing selected key developmental events.

Recently reported results of proximity placement analyses of human-specific NANOG-binding sites (HSNBS) revealed their associations with coding genes governing physiological development and functions of nervous and cardiovascular systems, embryonic development, behavior, as well as development of a diverse spectrum of pathological conditions (Glinsky, 2015). Here this approach was applied to dissect the expression profiles of protein-coding genes implicated in the development of the fetal and adult brain of *H.sapiens* (Zhang et al., 2011) and examine their co-localization patterns with HSNBS in the human genome. During the first step of this analysis, a set of 251 genes that are at least 20-fold up- or down-regulated in the neocortex temporal lobe of the fetal versus adult human brain has been identified. This set of 251 genes was designated a gene expression signature of the human fetal neocortex temporal lobe. It was of interest to ascertain the expression changes of these genes, if any, in the neocortex, the newest part of human brain, and in the prefrontal complex, which is implicated in complex cognitive behavior, compared with non-neocortical regions. Expression of two-third of genes comprising the 251-gene expression signature of the human fetal neocortex temporal lobe is significantly different in the neocortex compared with non-neocortical regions of human brain and gene expression profiles of these genes in neocortex and prefrontal cortex are highly concordant (**Figure 12**). Of note, 78% of genes differentially regulated in the neocortex versus non-neocortical regions manifest significant changes of expression in the prefrontal cortex versus non-neocortical regions of human brain as well. Remarkably, 92.6% of genes differentially regulated in both fetal versus adult brain and neocortex versus non-neocortical regions exhibit the same direction of expression changes, that is genes that are up-regulated in the fetal brain remain up-regulated in the neocortex and genes that are down-regulated in the fetal brain remain down-regulated in the neocortex (**Figure 12**). Thus, a majority of genes (60%) that acquired most dramatic expression changes in the human fetal neocortex temporal lobe appears to retain for many years the significance of expression changes acquired in the embryo and maintain highly concordant expression profiles in the neocortex and prefrontal cortex regions of adult human brain. Investigation of the placement enrichment pattern of HSNBS located near these neocortex/prefrontal cortex-associated genes revealed the most significant enrichment of HSNBS placement at the genomic distances less than 1.5 Mb with a sharp peak of the enrichment p value at the distance of 1.5 Mb (**Supplemental Figure S7**), suggesting that HSNBS may play a role in regulation of expression of these genes during embryogenesis.

These observations prompted the detailed analysis of all 12,885 genes, expression of which is significantly different in the human fetal neocortex temporal lobe compared with the adult brain ((Zhang et al., 2011), and examine their expression in the neocortex and prefrontal cortex regions of the adult human brain. Remarkably, the analysis of these fetal neocortex temporal lobe-associate genes revealed that expression of 4,958 genes is significantly different in the neocortex compared with non-neocortical regions of adult human brain and gene expression profiles of these genes in neocortex and prefrontal cortex are highly concordant (**Figure 12**). These results demonstrate that a very large number of genes, consisting of ~25% of all protein-coding genes in the human genome, acquires in embryo significant expression changes, which are stably maintained throughout the embryogenesis, infancy, and the adulthood as a highly concordant gene expression profile in the human fetal neocortex temporal lobe as well as in the neocortex and prefrontal cortex regions of adult human brain. It provides a compelling argument that gene expression profiles characteristic of the critical brain regions, which are essential for development of unique to human cognitive and behavioral functions, are established to a significant degree during the embryogenesis, retained for many years of human development, and maintained in the adulthood. In agreement with this idea, there is an increased enrichment of 5hmC at intragenic regions that are already hyper-hydroxymethylated at the fetal stage during human frontal cortex development, demonstrating that adult patterns of genic 5hmC in the frontal cortex are already evident in the immature fetal brain (Lister et al., 2013). In the human brain, transcriptional activity is associated with intragenic 5hmC enrichment and adult 5hmC patterns for cell type-specific genes appear established *in utero*; in contrast, loss of 5hmC enrichment is associated with developmentally coupled transcriptional down regulation of gene expression (Lister et al., 2013).

Recent experiments discovered that that there is a strikingly accelerated recruitment of new, evolutionary young genes during the early development of human brain leading to the increased expression of a significantly larger proportion of young genes in fetal or infant brains of humans compared with mouse (Zhang et al., 2011). Next, the detailed analysis of the evolutionary age of genes comprising the 4,958-gene expression signature of the neocortex/prefrontal cortex regions of human brain was carried-out. To this end, all 4,958 genes were segregated into thirteen sub-groups based on their respective evolutionary age, ranging from 0 (oldest genes) to 12 (youngest genes) as previously defined by Zhang et al. (2011). The gene expression enrichment factors were calculated for each individual evolutionary age sub-group by comparisons of corresponding gene age-associated distribution metrics, which were derived from the analysis of gene age-associated distribution profiles of all 19,335 genes interrogated in gene expression profiling experiments and 12,885 genes with expression changes significantly different in fetal versus adult brain. The resulting values for each evolutionary age sub-group of genes were normalized to the corresponding numerical values obtained for all genes within the corresponding set. The results of these analyses revealed that the relative enrichment factors of evolutionary young genes are higher compared to evolutionary older counterparts, which appears to be associated exclusively with the fraction of up-regulated genes of the 4,958-gene signature of the neocortex/prefrontal cortex regions of human brain (**Figure 12**).

Next, the placement enrichment analysis of HSNBS located in close proximity to the genomic coordinates of the 4,958 genes was performed. In these experiments, the estimates of two different placement enrichment metrics were computed. The numerical values of the first metric is based on the quantitation of genes residing within the genomic distance of 1.5 Mb or less from the nearest HSNBS (**Figure 12**). The threshold of 1.5 Mb was chosen based on the quantitative definition of the genomic distance associated with the most statistically significant enrichment of HSNBS placement near neocortex/prefrontal cortex associated genes (**Supplemental Figure S7**). The numerical values of the second metric is based on the quantitation of genes residing within and outside of the common TAD with HSNBS (**Figure 12**). In both experimental settings, the corresponding placement enrichment values for each sub-group of genes were compared to the values calculated for two control gene sets: 447 genes that are not expressed in the human brain and 357 expressed unbiased genes, i.e., genes that are uniformly expressed in human brain and do not manifest statistically significant differences of expression. Results of these analyses were highly consistent regardless of the utilized placement enrichment metrics: it appears that placements of HSNBS favors the evolutionary younger genes of sub-groups 7-12 compared with the evolutionary older genes or genes that are not expressed in the human brain (**Figure 12**). It is important to mention that all genes comprising both control gene sets were assigned to the evolutionary age groups 7-12, thus excluding the possibility that placement enrichment differences are due to genes of the control gene sets being belong to the evolutionary older genes.

Collectively, these data are highly consistent with the idea that HSNBS may have contributed to the establishment during the embryogenesis of the genome-wide expression changes characteristic of neocortex and prefrontal cortex regions of human brain, which are retained as a highly concordant 4,958-gene expression signature for many years of human development and maintained in the adulthood. The validity of this idea was further explored utilizing the Ingenuity Pathway Analysis software (http://www.ingenuity.com/ ) to identify and analyze possible developmental and pathophysiological regulatory networks comprising of HSNBS-associated genes, which appear implicated in development of human neocortex and prefrontal cortex regions. The Ingenuity software identified two main candidate regulatory networks and predicted five potential top regulators and eight their immediate target genes (**Figure 13**). Of note, seven of eight immediate target genes of the top five putative upstream regulators were also identified as regulatory elements of the two main networks, which are marked by the five-pointed stars in the **Figure 13**. One exception is the *INSM1* and *NEUROD1* genes that appear to form an interconnected and highly biologically significant axis, which plays a crucial role in maintaining mature pancreatic □-cell functions via cooperating interactions of *INSM1*, *NEUROD1*, and *FOXA2* genes and combinatorial binding of INSM1, NEUROD1, and FOXA2 proteins to regulatory DNA sequences (Jia et al., 2015). Furthermore, the *Insm1* gene is required for proper differentiation of all types of endocrine cells in the anterior pituitary gland, including pituitary cells producing thyroid-stimulating hormone, follicle-stimulating hormone, melanocyte-stimulating hormone, adrenocorticotrope hormone, growth hormone and prolactin (Welcker et al., 2013). Because it has been previously demonstrated that *Insm1* gene is required for development and differentiation of endocrine cells in the pancreas, intestine and adrenal gland (Gierl et al., 2006; Wildner et al., 2008), it has been defined as the essential pan-endocrine transcription factor (Welcker et al. (2013).

One of the intriguing features seemingly connecting these networks is that some genes identified as potential targets within one network appear within another network as the regulatory hubs, which are marked as the blue five-pointed stars in the **Figure 13**. This observation suggests that these networks are interconnected by the positive feedback regulatory loops, which appear designed to support the sustained networks' activity. The largest network is governed by the NANOG and POU5F1 proteins, which are also identified as potential targets of the second regulatory network. The *VEGFC* and *KDR* genes appear as the main regulatory hubs of the second putative regulatory network, within which the *NANOG* and *POU5F1* are depicted as potential target genes (**Figure 13**). Conversely, the *VEGFC* and *KDR* genes were identified as the putative targets of the regulatory network governed by the NANOG and POU5F1 proteins (**Figure 13**).

Taken together, the results of this analysis support the hypothesis that during embryogenesis the NANOG and POU5F1 proteins initiate the gene expression changes characteristic of neocortex and prefrontal cortex regions of human brain, which are discernable as a highly concordant 4,958-gene expression signature. At the later stages of brain development and functioning during infancy and transition to adulthood, the VEGFC and its receptor KDR may be responsible, in part, for maintenance and sustained activity of the fetal gene expression program in the neocortex/prefrontal cortex regions. This conclusion opens attractive opportunities for targeted pharmacological interventions using small molecule agonists and/or antagonists of the VEGFC/KDR axis *in vivo*.

**Placement enrichment analysis of HSNBS near genes encoding the protein partners of the NANOG-centered protein interaction network in embryonic stem cells**

One of the key mechanisms by which NANOG delivers its critical regulatory functions in ESC is the formation of protein-protein complexes with numerous protein partners, which are designated collectively as the NANOG interactome (Wang et al, 2006; Wu et al, 2006; Liang et al, 2008; Costa et al, 2013; Gagliardi et al., 2013). Recent high-definition analysis of physically interacting proteins in ESC using an improved affinity purification protocol expanded the NANOG interactome to more than 130 proteins (Gagliardi et al., 2013), many of which are components of large multi-subunit complexes involved in chromatin remodeling. Placement enrichment analysis identifies 58 genes encoding protein-protein interaction partners of NANOG that are located in close proximity to HSNBS defined by the 1.5 Mb placement enrichment metric. Genetic components of the multi-subunit chromatin remodeling complex NuRD appear most significantly overrepresented among genes encoding NANOG-interacting proteins: genes encoding ten of twelve NuRD complex protein subunits (83%) are located in close proximity to HSNBS (**Figure 13**). Notably, 27 of 58 genes encoding protein partners of NANOG that are located near HSNBS have been previously identified as the components of the POU5F1 (OCT4)-centered protein interaction network in ESC (van den Berg et al., 2011). Significantly, the efficient association in the ESC nucleus of both POU5F1 (OCT4) and NANOG proteins with all twelve proteins comprising known subunits of the classical NuRD complex has been documented (van den Berg et al., 2011; Gagliardi et al., 2013), which indicates that association of both NANOG and POU5F1 proteins with classical NuRD complex may constitute an important mechanism of their biological activity in ESC.

Collectively, the result of these analyses strongly argue that genes encoding proteins that physically interact with both NANOG and POU5F1 in the ESC nucleus may represent functionally-relevant targets of the HSGRL-associated regulatory networks. Genetic components of the classical multi-subunit chromatin remodeling complex NuRD seem particularly relevant in this regard and they should constitute the primary candidates for the follow-up structural-functional analyses and mechanistic experiments aiming to dissect their roles in human brain development and functions. Consistent with this hypothesis, recent experiments directly implicated the NuRD chromatin remodeling complex in establishing synaptic connectivity in the rodent brain (Yamada et al., 2014) and in directing the timely and stable peripheral nerve myelination by Schwann cells in mice (Hung et al., 2012). Furthermore, significant age-associated decline of expression in dentate gyrus sub-region of the hippocampus of one of the NuRD complex subunits, RbAp48 (RBBP4), was mechanistically connected with the age-related memory loss in humans (Pavlopoulos et al., 2013).

**Discussion**

Genome-wide proximity placement analyses of 10,598 DNA sequences representing four distinct families of HSGRL identified a small fraction of TADs in the human genome, which manifest a statistically significant accumulation of HSGRL compared to the expected values estimated based on a random distribution model. This set of TADs, termed rapidly-evolving in humans TADs (revTADs), acquired the maximum enrichment levels of 16-fold for HARs; 17-fold for hESC-FHSRRs; 50-fold for HSTFBS; and 88-fold for DHS-FHSRRs ($p < 0.0001$ in all instances; **Table 1**). Follow-up analyses of sixty revTADs, which were defined based on the enrichment of HARs and HSTFBS, revealed that all reported to date HSGRL and human-specific epigenetic signatures associated with embryonic development appear significantly enriched within the revTAD boundaries (**Table 2**). One of the features of revTADs was revealed by the correlation analyses: placements of different types of HSTFBS manifest significant positive correlations, whereas placements of HARs and HSTFBS exhibit inverse correlation profiles (**Figure 1**). A majority of revTADs in human genome tend to harbor a small number of coding genes: 65% of revTADs contain five or less protein-coding genes (**Supplemental Figure S8**), among which eight revTADs harbor just one protein-coding gene and two revTADs contain only non-coding RNA genes. Several revTADs contain large clusters of functionally-related protein-coding genes: three revTADS on chr11 harbor 61 genes encoding olfactory receptors; three consecutively-spaced revTADs on chr19 harbor 31 genes encoding zinc finger proteins; and one revTAD on chr20 contains eight beta-defensins' genes. Overview of the RNAseq data documented two other features of revTADs: i) clearly discernable tissue-specific patterns of non-coding RNA expression (**Supplemental Figure S8**); and ii) prominent presence of non-coding RNA transcripts in human brain (**Supplemental Figure S9**).

Correlation screens of proximity placement patterns of different HSGRL residing within the revTADs revealed markedly distinct correlation profiles of individual members of HSGRL families and recombination rates within the host revTADs. These analyses readily distinguish placement patterns of HARs and HSTFBS: revTADs having high recombination scores tend to accumulate large numbers of HARs while revTADs with low recombination scores harbor high numbers of HSTFBS (**Figure 2**). Consistent with the requirements of DNA strand interactions and strand invasion for the recombination process, significant direct correlations were observed between recombination scores and intra-chromosomal contacts within revTADs (**Figure 2**). Therefore, placements of HARs and HSTFBS within revTADs appear associated with distinct molecular processes. Placement of HARs within revTADs exhibits significant positive correlation with high recombination rates (**Figure 2**), which seems to connect the biogenesis of HARs with recombination mechanisms. In contrast, significant inverse correlation between HSTFBS placement and recombination rates as well as location of 99% of HSTFBS within TE – derived DNA sequences (Glinsky, 2015) strongly implicate TE activity in the biogenesis of HSTFBS. Since HARs by definition are located within highly evolutionary conserved DNA sequences, results of these analyses suggest that molecular mechanisms driving the emergence of regulatory loci evolved by exaptation of ancestral DNA (Cotney et al., 2013; Villar et al., 2015) may be associated with meiotic recombination as well.

One of the notable features of potential mechanistic significance is the association of hESC-enriched enhancers and SEs with HSGRL (**Table 3**). Taking into account that there are ~4,000 HSTFBS in the hESC genome (Glinsky et al., 2015) and nearly 60% of HARs overlap at least one of the common markers of enhancers in human cells (Capra et al., 2013), it seems logical to propose that creation of new enhancer elements leading to increasing density of conventional enhancers in selected genomic regions is one of the key events defining the increasing genomic complexity as the main direction of evolution of GRNs. Consistent with this idea, there are significant direct correlations between the placements within TADs of HARs and hESC-enriched enhancers (**Figure 3**), indicating that there is an apparent trend of the placement preference of HARs within TADs harboring hESC-enriched enhances. It follows, that higher density of conventional enhancers would increase the probability of local structural transitions of chromatin architecture to SEs and SEDs, provided other local structural requirements are in place or co-evolved (see below). In agreement with this model, there are 3-fold more SEs in the hESC genome compared with the mESC and the median size of hESC SEs is significantly larger (**Figure 4**).

Concomitantly, marked changes of TAD structural features in the hESC genome compared with mESC were observed, which are particularly striking for revTADs. Genome-wide, there are 42% more TADs in the hESC genome and the median size of hESC TADs is significantly smaller (**Figure 4**). These changes of TAD structural features remain consistent when the corresponding comparisons were made between the similar in size individual human and mouse chromosomes (**Figure 4**), indicating that hESC genome contains the increased number of the predominantly smaller size TADs. Consistent with these observations, comparisons of the high-resolution 3D maps of human and mouse genomes, which were recently obtained by Rao et al. (2014) using in situ Hi-C at a resolution range of 1 – 5 Kb, revealed that there are ~3-fold more contact domains formed by the concomitantly increased numbers of long-range chromatin loops in the human GM12878 B-cell lymphoblasts compared with the mouse CH12 B-cell lymphoblasts (Rao et al., 2014).

In the hESC genome, structural changes of the revTADs seem particularly evident: 32% of revTADs are located within primate-specific genomic regions and the remaining revTADs appear evolved via mechanisms of boundary crossing, domain mergers, and creation of new boundaries within larger TADs of orthologous mouse sequences. Collectively, this dramatic changes of the regulatory infrastructure of interphase chromatin can be explained by the model of convergence of TAD and SED architecture (**Figure 6**). According to this model, the high density of conventional enhancers increases the likelihood of transition to larger in size SE structures, formation of new SEDs, and increasing segmentation of genomic regions into insulated regulatory neighborhoods of large SED/small TAD structures. Consistent with this model, significant correlations were observed between the size of TADs and genomic span of hESC SEs residing within TADs (**Figure 5A**). There are significant direct correlations between the revTAD sizes and the numbers of hESC-enriched enhancers residing within revTADs (**Figure 5B**). Genome-wide, the highly significant direct correlation was observed between the size of TADs and the number of hESC-enriched enhancers located within TADs (**Figure 5C**).

In addition to enabling the creation of new enhancer elements, increasing density of conventional enhancers, and facilitating transition to SE structures, potential mechanisms of HSGRL-mediated effects on principal regulatory structures of interphase chromatin are likely involve emergence of overlapping CTCF/cohesin sites and LMNB1-binding sites as well as continuing insertion of Alu clusters near the putative DNA bending sites (**Figures 6 & 7**). Collectively, the ensemble of these structural changes facilitated by the targeted placements and retention of HSGRL at specific genomic locations would enable the emergence of new SED structures and remodeling of existing TADs to drive evolution of GRNs (**Figure 6**). Presented models of dynamic transitions of the interphase chromatin principal regulatory structures are in accord with the results of recent high-resolution in situ Hi-C experiments demonstrating that human genomes are partitioned into contact domains consisting of ~10,000 loops accommodating functional links between enhancers and promoters (Rao et al., 2014). Consistent with the idea that chromatin loops frequently demarcate the boundaries of contact domains, anchors at the loop bases typically occur at the contact domain boundaries and involve binding of two CTCF/cohesin sites in a convergent orientation with the asymmetric binding motifs of interacting sites aligned to face each other (Rao et al., 2014).

Compelling evidence of conservation in individual human genomes of different families of HSGRL (Marnetto et al., 2014; Glinsky, 2015; **Figure 8; Supplemental Figure S4**) are in accord with the hypothesis of their putative functional role in defining human-specific phenotypes. Observations of highly consistent patterns of significant direct correlations between conservation profiles of distinct seemingly unrelated HSGRL sequences in individual human genomes were observed (**Figure 8C** and **Supplemental Figure S4**) suggest that HSGRL conservation patterns reflect intrinsic features of the individual human genome, integration of which can help to assess the overall capacity of a genome to create and retain HSGRL.

The potential practical utility of the outlined here concepts are illustrated by building the working models of *NANOG*, *POU5F1*, and *POU3F2* super-enhancers' domains and associated TAD structures in the hESC genome (**Figures 9-11; Supplemental Figures S5, S6**). Combination of the utility functions of the UCSC Genome Browser and GEM principles (**Figure 6**) facilitated visualization of key structural elements of potential functional significance within each model of putative SED structures (**Figures 7; 9-10; Supplemental Figures S5; S6**). Utilizing structural-functional guidance derived from these models, conserved profiles and uniquely distinct patterns of H3K27ac peaks' distribution within the *NANOG* locus-associated genomic region in 43 types of human cells and tissues were inferred (**Figure 11**).

Putative regulatory consequences of targeted placements of HSGRL on expression of protein-coding genes were assessed during the analysis of expression profiles of protein-coding genes implicated in the development of the fetal and adult brain (Zhang et al., 2011). These analyses indicate that HSNBS may have contributed to the establishment during the embryogenesis of the genome-wide expression changes characteristic of neocortex and prefrontal cortex regions of human brain, which are retained as a highly concordant 4,958-gene expression signature for many years of human development and maintained throughout the adulthood (**Figure 12**). The follow-up Ingenuity pathway analysis confirmed that NANOG and POU5F1 proteins are most likely candidates to initiate the gene expression changes characteristic of neocortex and prefrontal cortex regions of human brain (**Figure 13**), which are discernable as a highly concordant 4,958-gene expression signature. At the later stages of brain development and function during infancy and transition to adulthood, the VEGFC and its receptor KDR may be responsible, in part, for maintenance and sustained activity of the fetal gene expression program in the neocortex/prefrontal cortex regions.

Strikingly, placement enrichment analysis of HSNBS near genes encoding the protein partners of the NANOG-centered protein interaction network in embryonic stem cells identified ten genes encoding protein subunits of the classical NuRD multi-subunit chromatin remodeling complex as the candidate principal genomic regulatory elements, activity of which appears preferentially targeted by insertions of HSRGL. Importantly, all these regulatory proteins were previously identified as the protein-protein interactions partners of both NANOG and POU5F1 (OCT4) proteins in the hESC nucleus (van den Berg et al., 2011; Gagliardi et al., 2013), thus, supporting the idea that activities of NANOG and POU5F1 proteins via engagement of the classical NuRD multi-subunit chromatin remodeling complex play a central role in defining key human-specific elements of gene expression changes during the embryonic development of human brain.

**Concluding remarks**

Chromosomes can be viewed as the physical conduits of genetic information enabling its secure storage, maintenance, transfer, and efficient translation into a diverse spectrum of specific phenotypes. To a large degree all these processes are facilitated by the linear code of DNA sequences and enabled by the chromatin structures defining a 3D architecture of interphase chromosomes, which can be considered collectively as the chromatin folding code. Present analyses indicate that increasing regulatory complexity in human genomes associated with targeted placements of thousands HSGRL has major effect on the principal regulatory structures of interphase chromatin, namely TADs, SEs, and SEDs. One of the principal creative events in this continuing chromatin domain architecture remodeling process is the HSGRL-enabled emergence of new enhancer elements, increasing density of which would increase the probability of structural transition from conventional enhancers to novel SEs and formation of new SEDs. Significantly, presence of conserved and/or newly created overlapping CTCF/cohesin-binding sites flanking novel SEs is the essential requirement for the formation of functional SEDs. This continuous process involves continuing removal of old and creation of new TFBS through multiple trial-and-error events enabled by retrotransposition, cytosine MADE, and recombination. Furthermore, present analyses suggest that these HSGRL-associated changes of the nuclear regulatory architecture may facilitate the enhanced precision of regulatory interactions between enhancers and target genes within reconstructed insulated SED neighborhoods and rewired TAD networks. Taken together, these observations imply that convergence of TAD and SED architectures, which is exemplified by concomitant processes of increasing both quantity and size of SEDs and the increasing number and size reduction of TADs, is one of the main directions of the interphase chromatin structural changes during the structural evolution of genomic regulatory networks.

Comparisons of estimates of time periods required for creation of enhancers and SEs (**Table 6**) with recently reported estimates of enhancers' half-lives and mean-lifetimes (Villar et al., 2015), which were estimated at 296 and 427 hundred million years, respectively, seem to indicate that new enhancer elements are created at a markedly faster pace during evolution compared with their decay time span. This marked dichotomy of time period requirements for new enhancers' creation versus enhancers' decay provides an underlying mechanism for increasing genomic complexity as a major trend during evolution of genomic regulatory networks. A vast majority of distinct classes of regulatory elements in the hESC genome appears created on conserved DNA sequences (**Table 6**; **Supplemental Figure S3**), suggesting that exaptation of ancestral DNA constitutes a main mechanism of creation of new regulatory sequences in the human genome. This conclusion is highly congruent with recent reports describing exaptation of ancestral DNA as a mechanism of creation of human-specific enhancers active in embryonic limb (Cotney et al., 2013) and as a prevalent mechanism of recently evolved enhancers' creation during the mammalian genome evolution (Villar et al., 2015). Taken together with results of the present analyses, these observations seem to support the hypothesis that meiotic recombination is the predominant mechanism responsible for creation of new regulatory loci, since exaptation of ancestral DNA appears to generate overall many more recently evolved regulatory elements than can be attributed to the repeat-driven expansion mechanisms.

It has been suggested that evolution of enhancers may represent an example of evolution of evolvability (Duque and Sinha, 2012), according to which a genotypic feature evolves not due to its functional effect, but due to its effect on the ability of DNA to evolve more quickly (Wagner and Altenberg, 1996). In this context, the regulatory balance of HAT and HDAC activities at specific genomic locations as well as the performance of enzymatic systems regulating cytosine recovery/methyl-cytosine deamination cycles and DNA recombination processes may play a major role during the evolution of regulatory DNA sequences and emergence of GRNs controlling unique to human phenotypes. Critical experimental and clinical assessments of potential therapeutic opportunities for pharmacological interventions targeting these enzymatic systems and functions of VEGFC/KDR axis would be of immediate interest.

**Materials and Methods**

*Data Sources and Analytical Protocols*

Solely publicly available datasets and resources were used for this analysis as well as methodological approaches and a computational pipeline validated for discovery of primate-specific gene and human-specific regulatory loci (Tay et al., 2009; Kent, 2002; Schwartz et al., 2003; Capra et al., 2013; Marnetto et al., 2014; Glinsky, 2015). The analysis is based on the University of California Santa Cruz (UCSC) LiftOver conversion of the coordinates of human blocks to corresponding non-human genomes using chain files of pre-computed whole-genome BLASTZ alignments with a minMatch of 0.95 and other search parameters in default setting (http://genome.ucsc.edu/cgi-bin/hgLiftOver). Extraction of BLASTZ alignments by the LiftOver algorithm for a human query generates a LiftOver output "Deleted in new", which indicates that a human sequence does not intersect with any chains in a given non-human genome. This indicates the absence of the query sequence in the subject genome and was used to infer the presence or absence of the human sequence in the non-human reference genome. Human-specific regulatory sequences were manually curated to validate their identities and genomic features using a BLAST algorithm and the latest releases of the corresponding reference genome databases for time periods between April, 2013 and June, 2015.

Genomic coordinates of 3,127 topologically-associating domains (TADs) in hESC; 6,823 hESC-enriched enhancers; 6,322 conventional and 684 super-enhancers (SEs) in hESC; 231 SEs and 197 SEDs in mESC were reported in the previously published contributions (Dixon et al., 2012; Xie et al., 2013; Hnisz et al., 2013; Whyte et al., 2013; Dowen et al., 2014). The primary inclusion criterion for selection of the human-specific genomic regulatory loci (HSGRL) analyzed in this contribution was the fact that they were identified in human cells lines and primary human tissues whose karyotype were defined as "normal". The following four HSGRL families comprising of 10,598 individual regulatory DNA sequences were analyzed in this study: 1) Human accelerated regions (HARs; Capra et al., 2013); 2) Human-specific transcription factor-binding sites (HSTFBS; Glinsky, 2015); 3) hESC-derived fixed human-specific regulatory regions (hESC-FHSRR; Marnetto et al., 2014); 4) DNase hypersensitive sites-derived fixed human-specific regulatory regions (DHS-FHSRR; Marnetto et al., 2014). The number of HSGRL placed within a given TAD was computed for every TAD in the hESC genome and the HSGRL placement enrichment was calculated as the ratio of observed values to expected values estimated from a random distribution model at the various cut-off thresholds. Datasets of NANOG-, POU5F1-, and CTCF-binding sites and human-specific TFBS in hESCs were reported previously (Kunarso et al., 2010; Glinsky, 2015) and are publicly available. RNA-Seq datasets were retrieved from the UCSC data repository site (http://genome.ucsc.edu/; Meyer et al., 2013) for visualization and analysis of cell type-specific transcriptional activity of defined genomic regions. A genome-wide map of the human methylome at single-base resolution was reported previously (Lister et al., 2009; 2013) and is publicly available (http://neomorph.salk.edu/human_methylome). The histone modification and transcription factor chromatin immunoprecipitation sequence (ChIP-Seq) datasets for visualization and analysis were obtained from the UCSC data repository site (http://genome.ucsc.edu/; Rosenbloom et al., 2013). Genomic coordinates of the RNA polymerase II (PII)-binding sites, determined by the chromatin integration analysis with paired end-tag sequencing (ChIA-PET) method, were obtained from the saturated libraries constructed for the MCF7 and K562 human cell lines (Li et al., 2012). Genome-wide maps of interactions with nuclear lamina, defining genomic coordinates of human and mouse lamin-associated domains (LADs), were obtained from previously published and publicly available sources (Guelen et al., 2008; Peric-Hupkes et al., 2010). The density of TF-binding to a given segment of chromosomes was estimated by quantifying the number of protein-specific binding events per 1-Mb and 1-kb consecutive segments of selected human chromosomes and plotting the resulting binding site density distributions for visualization. Visualization of multiple sequence alignments was performed using the WebLogo algorithm (http://weblogo.berkeley.edu/logo.cgi ). Consensus TF-binding site motif logos were previously reported (Kunarso et al., 2010; Wang et al., 2012; Ernst and Kellis, 2013).

      The quantitative limits of proximity during the proximity placement analyses were defined based on several metrics. One of the metrics was defined using the genomic coordinates placing HSGRL closer to putative target protein-coding or lncRNA genes than experimentally defined distances to the nearest targets of 50% of the regulatory proteins analyzed in hESCs (Guttman et al., 2011). For each gene of interest, specific HSGRL were identified and tabulated with a genomic distance between HSGRL and a putative target gene that is smaller than the mean value of distances to the nearest target genes regulated by the protein-coding TFs in hESCs. The corresponding mean values for protein-coding and lncRNA target genes were calculated based on distances to the nearest target genes for TFs in hESC reported by Guttman et al. (2011). In addition, the

proximity placement metrics were defined based on co-localization within the boundaries of the same TADs and the placement enrichment pattern of HSNBS located near the 251 neocortex/prefrontal cortex-associated genes, which identified the most significant enrichment of HSNBS placement at the genomic distances less than 1.5 Mb with a sharp peak of the enrichment p value at the distance of 1.5 Mb (**Supplemental Figure S7**).

The comprehensive database of expression profiles of protein-coding genes implicated in the development of the fetal and adult brain of *H. sapiens* was obtained from the previously published contribution (Zhang et al., 2011). Analysis of the evolutionary age of genes comprising the 4,958-gene expression signature of the neocortex/prefrontal cortex regions of human brain was carried-out by segregating genes into thirteen sub-groups based on their respective evolutionary age, ranging from 0 (oldest genes) to 12 (youngest genes) as previously defined by Zhang et al. (2011). The gene expression enrichment factors were calculated for each individual evolutionary age sub-group by comparisons of corresponding gene age-associated distribution metrics, which were derived from the analysis of gene age-associated distribution profiles of all 19,335 genes interrogated in gene expression profiling experiments and 12,885 genes with expression changes significantly different in fetal versus adult brain. The resulting values for each evolutionary age sub-group of genes were normalized to the corresponding numerical values obtained for all genes within the corresponding set.

The assessment of conservation of HSGRL in individual genomes of 3 Neanderthals, 12 Modern Humans, and the 41,000-year old Denisovan genome (Reich et al., 2010; Meyer et al., 2012) was carried-out by direct comparisons of corresponding sequences retrieved from individual genomes and the human genome reference database (http://genome.ucsc.edu/Neandertal/ ). Direct access to the specific Genome Browser tracks utilized for analyses and visualization:

http://genome.ucsc.edu/cgi-bin/hgTracks?db=hg18&position=chr10%3A69713986-69714099&hgsid=393865029_yg7UixUE4a4awjjTahns4KTPkII1 .

Recombination rates were downloaded from the HapMap Project (The International Hapmap Consortium, 2007) and the numbers of DNA segments with the recombination rates of 10 cM/Mb or greater were counted. This threshold exceeds ~10-fold the mean intensity of recombination rates in telomeric regions, which were identified as the regions with the higher recombination rates in the human genome. It is well known that over large genomic scales, recombination rates tend to be higher in telomeric as compared to centromeric chromosomal regions. In telomeric regions, the mean detected hotspot spacing is 90 kb and the mean intensity (total rate across the hotspot) per hotspot is 0.115 cM, whereas for centromeric regions the mean spacing is 123 kb and the mean intensity is 0.070 cM (The International Hapmap Consortium, 2007).

*Statistical Analyses of the Publicly Available Datasets*

All statistical analyses of the publicly available genomic datasets, including error rate estimates, background and technical noise measurements and filtering, feature peak calling, feature selection, assignments of genomic coordinates to the corresponding builds of the reference human genome, and data visualization, were performed exactly as reported in the original publications and associated references linked to the corresponding data visualization tracks (http://genome.ucsc.edu/). Any modifications or new elements of statistical analyses are described in the corresponding sections of the Results. Statistical significance of the Pearson correlation coefficients was determined using GraphPad Prism version 6.00 software. The significance of the differences in the numbers of events between the groups was calculated using two-sided Fisher's exact and Chi-square test, and the significance of the overlap between the events was determined using the hypergeometric distribution test (Tavazoie et al., 1999).

**Supplemental Information**

Supplemental information includes Supplemental Table S1, Supplemental Figures S1–S8 and can be found with this article online.

**Author Contributions**

This is a single author contribution. All elements of this work, including the conception of ideas, formulation, and development of concepts, execution of experiments, analysis of data, and writing of the paper, were performed by the author.

**Acknowledgements**

This work was made possible by the open public access policies of major grant funding agencies and international genomic databases and the willingness of many investigators worldwide to share their primary research data. I would like to thank my colleagues for their valuable critical contributions during the informal review and formal peer review process of this work.

**Figure Legends**

**Figure 1.** Correlation screens revealed distinct patterns of associations between HARs and TFBS residing within the rapidly-evolving in human Topologically Associating Domains (revTADs).

A. UCSC Genome Browser view of the revTAD on human chr6 harboring 10 Human Accelerated Regions, HARs (red bars), 10 hESC-enriched enhancers (black bars), 52 primate-specific TFBS for NANOG (26 sites), POU5F1 (10 sites), CTCF (26 sites), and 72 recombination hotspots with recombination rates at least 10 cM/Mb (blue bars). Genomic coordinates of POU3F2 super-enhancer domain in the hESC genome is depicted by the horizontal arrow.

B. Direct correlations between the numbers of primate-specific CTCF-binding sites and HSTFBS residing within the 60 revTADs (top left panel). Note that when primate-specific CTCF-binding sites were segregated into human-specific and non-human primate specific sub-groups, correlation coefficients are stronger for human-specific CTCF-binding sites (top right and bottom left panels) and weaker for non-human primate-specific CTCF-binding sites (bottom right panel).

C. Inverse correlations between the numbers of primate-specific CTCF-binding sites and HARs residing within the 60 revTADs (top left panel). Note that when primate-specific CTCF-binding sites were segregated into human-specific and non-human primate specific sub-groups, correlation coefficients are stronger for human-specific CTCF-binding sites (top right panel) and weaker for non-human primate-specific CTCF-binding sites (bottom right panel). Bottom left panel shows the inverse correlation between the numbers of human-specific NANOG-binding sites and HARs.

D. The inverse correlation between the numbers of HSTFBS and HARs residing within the 36 revTADs each of which contains both HSTFBS and HARs (top left panel). Top right panel shows the inverse correlation between the numbers of HSTFBS and HARs residing within the set of 60 revTADs (see Table 2 and text for details).

**Figure 2.** Distinct correlation profiles of HSGRL and recombination rates within the revTADs distinguish placement patterns of HARs and HSTFBS.

A, B. Visualization of the placement distribution patterns of HARs (low positioned red bars), recombination hotspots, RHs (blue bars), and HSTFBS (high positioned red bars with designations of TF names) within the revTADs. Note that revTADs containing high numbers of RHs (192 to 402 RHs) tend to harbor higher numbers of HARs (15 to 24 HARs) and no HSTFBS (figures in the panel A). In contrast, revTADs containing intermediate (69 to 94 RHs) or low (2 RHs) numbers of RHs tend to harbor intermediate and low numbers of HARs and multiple HSTFBS (figures in the panel B).

C. Inverse correlations between the numbers of RHs (designated as the recombination scores) and numbers of HSTFBS within the revTADs. Correlation profiles are shown for recombination scores (RS) and HSTFBS (top left panel), RS and HSNBS (top right panel), and RS and human-specific CTCF-binding sites (bottom left panel). In contrast, lack of significant correlation was recorded for RS and the numbers of non-human primate-specific CTCF-binding sites (bottom right panel).

D. Significant direct correlations were observed between the RS and the numbers of HARs (top left panel) and the RS and cumulative numbers of intrachromosomal contacts (top left panel) within the revTADs. Bottom two panels show the correlation profiles between the median RS values and median values of intrachromosomal contacts (bottom left panel) and the median numbers of HARs and HSTFBS (bottom left panel) for the four sub-groups of revTADs that were segregated into quartiles based on the values of the cumulative numbers of intrachromosomal contacts observed within the individual revTADs (Jin et al., 2013). Note that the median RS value for the revTADs placed in the top quartile based on the quantity of observed intrachromosomal interactions was 40-fold greater compared to the median RS value of the revTADs assigned to the bottom quartile (bottom left panel).

**Figure 3.** Genome-wide associations between the number of HARs and the quantity of hESC-enriched enhancers located within TADs.

A. Direct correlation between the numbers of HARs and hESC-enriched enhancers located within the individual revTADs.

B-D. In the hESC genome, TADs segregated into sub-groups with increasing numbers of HARs contain concomitantly higher proportion of hESC-enriched enhancer-harboring TADs compared with TADs without hESC enhancers. The fractions of TADs with and without hESC enhancers are shown for a total of 3,062 TADs (hg19 release of human reference genome database) in the hESC genome (B), for 1,135 TADs harboring 2,609 HARs (C), for 2,075 TADs containing 6,703 hESC-enriched enhancers (D). In the figure (D) the average numbers of hESC-enriched enhancers are shown for sub-groups of TADs harboring 0; 1; 2; and at least 5 HARs. P values designate the statistical significance of the differences of the hESC enhancer numbers between the neighboring sub-groups of TADs. Calculations of the HAR numbers located within TADs were performed after converting the genomic coordinates of 3,127 TADs (Dixon et al., 2012) from the hg18 to hg19 release of the human reference genome database using the LiftOver algorithm.

**Figure 4.** Markedly distinct structural features of SEs and TADs in genomes of human and mouse ESCs.

A. Size distribution analyses revealed significantly increased size of SEs and decreased size of TADs in the hESC genome compared with mouse. Top two panels show the median sizes of SEs (top left panel) and TADs (top right panel) sorted in a descending order based on individual TAD sizes and segregated into ten sub-groups at 10% increments. Bottom two panels show the median (bottom left panel) and average (bottom right panel) sizes of TADs for individual chromosomes of human and mouse ESC genomes. Note that median sizes of TADs on 17 of 20 (85%) chromosomes in the hESC genome (except chr7, chr13, and chr14) are smaller compared to median TAD sizes on chromosomes in the mESC genome.

B. Numbers of both SEs (top left panel) and TADs (top right panel) are increased in the hESC genome compared with mouse. Bottom two panels illustrate that numbers of TADs are decreased in differentiated cells compared to ESC both in humans (bottom left panel) and mice (bottom right panel).

**Figure 5.** Placements of hESC enhancers and HARs within TADs are directly correlated with the size of TADs in the hESC genome.

A. Significant direct correlations between the size of TADs and SE's span defined as a number of bp between the two most distant SEs located within a given TAD in the hESC genome (top two panels). Top left panel shows a correlation profile between the size of 504 TADs and genomic span of 642 SEs located within TADs. Top right panel shows a correlation profile between the size of 103 TADs harboring at least two SEs and genomic span of 241 SEs residing within TADs. Percentiles within shaded areas indicate the percent of TADs containing HSGRL within a designated set, which increases concomitantly with the increasing quantity of SEs located within TADs and larger TAD size. Note that similar trends were observed in the mESC genome, however, the correlation coefficient values were not statistically significant (bottom two panels).

B. Patterns of significant direct correlations between the size of 147 revTADs and numbers of hESC-enriched enhancers (top left panel) and HARs (top right panel) located within the revTADs. Bottom two panels show the profiles of significant correlations between residing within the revTADs hESC-enriched enhancers and HARs (direct correlation; bottom left panel) and HSTFBS and HARs (inverse correlation; bottom right panel).

C. Genome-wide, there is a highly significant direct correlation between the numbers of hESC-enriched enhancers located within TADs and the average size of corresponding TADs harboring hESC-enriched enhancers. The numbers of hESC-enriched enhancers located within each individual TAD were quantified and TADs harboring 0 to 9 enhancers were segregated into subgroups harboring the same numbers of enhancers. The numbers of TADs in each subgroup are indicated. Eighty-one TADs harboring ten or more hESC-enriched enhancers (range 10-30) were segregated into one subgroup with the average enhancers' content of 12.5 per TAD. The average TAD sizes were computed for each subgroup of TADs and the results were plotted to asses the correlation pattern.

**Figure 6.** A model of genome evolution driven by the increasing complexity of genomic regulatory networks (GRNs). It is proposed that evolutionary process occurs within the context of the intrinsic division of mammalian genomes into regions of high and low recombination rates that are associated with the low and high probabilities of TE insertion and/or retention as well as C/G and A/T alleles' bias, respectively. According to the genome evolution model (GEM), the continuing emergence of new enhancer elements constitutes a critical creative event driving the increasing complexity of GRNs in the hESC genome (see text for details). Potential mechanisms of HSGRL-mediated effects on principal regulatory structures of interphase chromatin involve: I) Creation of new TFBS and novel enhancer elements; II) Increasing density of conventional enhancers and facilitating transition to SE structures; III) Emergence of overlapping CTCF/cohesion-binding sites and LMNB1-binding sites;

IV) Continuing insertion of clusters of Alu elements near the putative DNA bending sites (**Figure 7**). Collectively, the ensemble of these structural changes facilitated by the targeted placements and retention of HSGRL at specific genomic locations would enable the emergence of new SED structures and remodeling of existing TADs to drive evolution of GRNs. MADE, cytosine methylation associated DNA editing.

**Figure 7.** Clusters of Alu elements in the vicinity of putative DNA bending sites near the borders of SEDs and TADs.

A. Clusters of Alu elements near the borders of the ID3 SED on chr1:23,878,033-23,894,300 (16,268 bp). Dotted lines depict the genomic positions of the overlapping CTCF/cohesin-binding sites, interactions of which form the anchor base of the ID3 SED.

B. Clusters of Alu elements near the NANOG SED left border on chr12:7,864,594-7,869,500 (4,907 bp).

C. Clusters of Alu elements near the NANOG SED right border on chr12:8,012,400-8,017,400 (5,001 bp).

Arrows in the figures (B) and )C) point the overlapping CTCF/cohesin-binding sites, interactions between which form the anchor base of the NANOG SED.

Note that clusters of closely-spaced sequences of at least three Alu elements belonging to most ancient AluJ (~65 million years old), second oldest AluS (~30 million years old), and currently active modern AluY sub-families are observed, suggesting that placement and/or retention of Alu elements at these sites were occurring for millions of years and continues at the present time (see text for details).

**Figure 8.** Conservation patterns of HSGRL in individual human genomes.

Conservation of HSGRL in individual genomes of 3 Neanderthals, 12 Modern Humans, and the 41,000-year old Denisovan genome was carried-out by direct comparisons of corresponding sequences retrieved from individual genomes and the human genome reference database (http://genome.ucsc.edu/Neandertal/). Full-length sequence alignments with no gaps of the individual human genome sequences to the corresponding sequences in the human genome reference databases were accepted as the evidence of sequence conservation in the individual human genome.

A. Top two panels show conservation patterns of HSGRL located within the revTADs for 57 HARs and associated 75 CTCF-binding sites (top left panel); 22 HARs harboring 23 LMNB1-binding sites and 23 LMNB1-binding sites residing within 22 HARs (top right panel). Bottom two panels show conservation patterns of 90

HARs harboring 93 LMB1-binding sites (bottom left panel) and 55 HARs harboring 55 CTCF-binding sites and 55 CTCF-binding sites located within 55 HARs (bottom right panel).

B. Top two panels show conservation patterns of 123 HARs associated with 127 high-confidence overlapping CTCF/RAD21-binding sites (top left pane) and 127 high-confidence overlapping CTCF/RAD21-binding sites associated with 123 HARs (top right panel). Bottom two panels show conservation patterns of 69 CTCF/RAD21-binding sites conserved in all Neandertals' genome (bottom left panel) and 152 human-specific NANOG-binding sites highly-conserved in individual human genomes (bottom right panel). Note that conservation of all HSGRL are consistently at the lowest level in the Neanderthals' genome, suggesting that creation and/or retention rates of HSGRL are enhanced in Modern Humans. However, HSGRL sequences manifesting the relatively high conservation levels in the Neanderthals' genome appear most conserved in the individual genomes of Modern Humans as well, including the 41,000-year old Denisovan genome (bottom left panel). Sequences of HSGRL with assigned biochemical functions, e.g., specific TFBS or Lamin B1 (LMNB1)-binding sites, which are residing within HARs exhibit markedly higher conservation levels compared to sequences of HARs harboring the corresponding HSGRL. This conclusion remains valid for HSGRL sequences with assigned specific biochemical or biological functions that were associated with HARs by proximity placement analyses (figures A, B).

C. Direct correlations of conservation profiles of DNA sequences of distinct HSGRL in individual human genomes (top two panels) and markedly different values of genomic fitness scores (GFS) integrating into a single numerical value sequence conservation data of 909 HSGRL in individual human genomes (bottom panel). Note that values of GFS are inversely correlated with the variation coefficients of GFS values in individual human genomes, consistent with the hypothesis that GFS reflects the intrinsic property of an individual genome to create and/or retain the HSGRL.

**Figure 9.** Genomic features of 2D models of individual super-enhancer domains (SEDs) in the hESC genome.

A. UCSC Genome Browser view of the 2D model of the NANOG super-enhancers' domain in the hESC genome (chr12:7,760,238-7,904,075). Genomic coordinates of two hESC-super-enhancers are depicted by the semi-transparent arks; positions of 5 primate-specific NANOG-binding sites are marked by the red arrows;

conserved and primate-specific overlapping CTCF/cohesin-binding sites that may be involved in formation of the anchors at the SED base are designated; genomic position of the primate-specific region harboring LAD sequence is depicted by the brace shape. Note that insertion of the primate-specific LAD sequence downstream of the NANOG gene suggests that SED loop is more likely to be formed by folding of the upstream chromatin chain toward the nuclear lamina-bound DNA sequence (figure 10).

B. UCSC Genome Browser view of the POU5F1 super-enhancer domain in the hESC genome (chr6:31,212,185-31,272,994). Genomic coordinates of the hESC-super-enhancer are depicted by the semi-transparent ark; positions of 3 primate-specific and one human-specific NANOG-binding sites are marked by the red arrows; conserved and primate-specific overlapping CTCF/cohesin-binding sites that may be involved in formation of the anchors at the SED base are designated.

C. UCSC Genome Browser zoom-in views of the anchor regions at the bases of SEDs are shown. Note that the SED anchor regions typically contain two LMNB1-binding sites near the overlapping CTCF/cohesin-binding sites, interactions of which secure the chromatin loop formation and segregation of insulated genomic neighborhoods of SEDs. These observations suggest that a spatial positioning of SED anchor regions is secured by the attachment to the nuclear lamina.

**Figure 10.** A snap-shot of principal genomic features of the 3D model of the interphase chromatin folding. A 3D model of spatial positioning of principal regulatory elements of the 2.54 Mb genomic region adjacent to the NANOG locus in the hESC genome is shown within the context of the 3 consecutive TADs and 4 boundary regions (depicted as blue clouds containing clusters of 3-5 overlapping CTCF/cohesin-binding sites). Positions of overlapping CTCF/cohesin-binding sites interactions of which secure the bases of chromatin loops are depicted by the yellow-colored donut shapes. Positions of key TFBS, including HSTFBS, are marked by the arrows. Attachment regions of blue-colored chromatin chains to the nuclear lamina are depicted by the placement of the corresponding segments within red-colored LADs. ED, hESC-enriched enhancers' domain; SED, super-enhancers' domains; EDs and SEDs are designated by the corresponding names of the protein-coding genes. Genomic coordinates of TADs, EDs, and SEDs are listed. Inset shows a highly significant direct

correlation between the size of TADs and genomic span of hESC-enriched enhancers located within TADs (r = 0.995; p = 0.0055).

**Figure 11.** Conserved profiles and uniquely distinct patterns of H3K27ac peaks' distribution within the NANOG locus-associated genomic region in 43 types of human cells and tissues.

Regional genomic maps of H3K27ac peaks/SE distributions depicting numbers of SEs placed within 3.26 Mb genomic region containing four TADs and three TAD boundaries (A) for individual types of human cells and tissues against the background of SE distribution profiles within the same region in the hESC genome. Note that all individual profiles of SE distributions in human cells and tissues (B, C), which reflect distribution profiles of H3K27ac peaks, follow the general pattern of SE distributions established in the hESC genome (figures B & C). One of the notable common elements that emerged in many differentiated cells and tissues is the appearance of SEs within the genomic regions demarcating TAD boundaries in the hESC genome, suggesting that TAD boundaries and TAD structures undergo visible rearrangements in differentiated cells and tissues. Erasing the hESC TAD boundaries during differentiation due to appearance of new cell-type specific SEs is consistent with observed decreasing numbers of TADs in differentiated cells and tissues (figure 4).

D, E. Unique patterns of H3K27peaks'/enhancers' distribution profiles in 43 types of human cells and tissues. In figure E, the data set was sorted in descending order of total numbers of enhancers (right Y axis) located within the region in each cell and tissue type. The position of hESC within this distribution is shown by the dotted line. The names of cells and tissues are listed. See text for details.

**Figure 12.** Identification and characterization of the 4,958 genes' expression signature in the human fetal neocortex temporal lobe retaining the expression changes acquired in the embryo for many years of human brain development and maintaining highly concordant expression profiles in the neocortex and prefrontal cortex regions of adult human brain.

A. Concordant expression profiles of 163 genes with at least 20-fold expression changes in the fetal neocortex temporal lobe that also manifest significant expression changes in the neocortex versus non-neocortical regions of adult human brain (top right panel). Note that expression changes of these 163 genes are highly concordant in both neocortex and prefrontal cortex regions of adult human brain versus non-neocortical regions (top left panel). Bottom two panels show highly concordant expression profiles of 4,958 genes with significant expression changes in fetal versus adult human brain and neocortex versus non-neocortical regions of adult human brain (bottom right panel) as well as either prefrontal cortex or neocortex versus non-neocortical regions of adult human brain (bottom left panel).

B. Analysis of expression enrichment patterns of 4,958 genes suggests a bias toward up-regulation of evolutionary young genes and down-regulation of evolutionary old genes. See text for detail.

C. Placement enrichment analysis of neocortex/prefrontal cortex signature genes near human-specific NANOG-binding sites (HSNBS) suggests a bias toward evolutionary young coding genes and genes expressed in brain compared with genes that are not expressed in human brain. Bottom right panel shows enrichment of co-localization within common TADs for neocortex/prefrontal cortex signature genes and HSNBS compared with genes that are not expressed in human brain.

**Figure 13.** Genomic regulatory networks (GRNs) of human neocortex/prefrontal cortex development and functions that are associated with HSNBS and under potential regulatory controls of the classical multi-subunit NurD chromatin remodeling complex.

A. Ingenuity Pathway Analysis of human neocortex/prefrontal cortex signature genes identifies candidate GRNs that appear under putative regulatory controls of the NANOG and POU5F1 proteins in embryogenesis and seem to transition to regulatory control of the VEGFC and its receptor KDR at the later developmental stages and in the adult brain. See text for details.

B. All 12 proteins comprising the classical multi-subunit NurD chromatin remodeling complex represent protein partners of both NANOG-centered and POU5F1-centered protein-protein interaction networks in hESC (van den Berg et al., 2011; Gagliardi et al., 2013). Genes encoding 10 of 12 (83%) subunits of the classical multi-subunit NurD chromatin remodeling complex are located near human-specific NANOG-binding sites in the hESC genome. See text for additional details.

**Table 1. Summary of search for rapidly-evolving topologically-associating domains (revTADs) in the hESC genome**

**1.1. Number of revTADs harboring at least 10 human-specific genomic regulatory loci (HSGRL) of the same HSGRL family**

| HSGRL family | HSGRL in human genome | revTAD number | Cumulative HSGRL number | Expected number | Fold enrichment | Percent TAD | Percent HSGRL | P value |
|---|---|---|---|---|---|---|---|---|
| HARs | 2745 | 21 | 286 | 18 | 16 | 0.7 | 10.4 | < 0.0001 |
| HSTFBS | 3803 | 33 | 1808 | 40 | 45 | 1.1 | 47.5 | < 0.0001 |
| hESC-FHSRR | 1932 | 6 | 62 | 4 | 17 | 0.2 | 3.2 | < 0.0001 |
| DHS-FHSRR | 2118 | 49 | 1440 | 33 | 43 | 1.6 | 68.0 | < 0.0001 |

**1.2. Number of revTADs harboring at least 40% of human-specific genomic regulatory loci (HSGRL) of the same HSGRL family**

| HSGRL family | HSGRL per one revTAD | revTAD number | Cumulative HSGRL number | Expected number | Fold enrichment | Percent TAD | Percent HSGRL | P value |
|---|---|---|---|---|---|---|---|---|
| HARs | 4 | 196 | 1223 | 172 | 7 | 6.3 | 44.6 | < 0.0001 |
| HSTFBS | 9 | 35 | 1826 | 43 | 43 | 1.1 | 48.0 | < 0.0001 |
| hESC-FHSRR | 4 | 161 | 857 | 99 | 9 | 5.1 | 44.4 | < 0.0001 |
| DHS-FHSRR | 29 | 16 | 949 | 11 | 88 | 0.5 | 44.8 | < 0.0001 |

**1.3. Number of revTADs harboring at least 50% of human-specific genomic regulatory loci (HSGRL) of the same HSGRL family**

| HSGRL family | HSGRL per one revTAD | revTAD number | Cumulative HSGRL number | Expected number | Fold enrichment | Percent TAD | Percent HSGRL | P value |
|---|---|---|---|---|---|---|---|---|
| HARs | 3 | 322 | 1601 | 283 | 6 | 10.3 | 58.3 | <0.0001 |
| HSTFBS | 5 | 53 | 1936 | 64 | 30 | 1.7 | 50.9 | <0.0001 |
| hESC-FHSRR | 3 | 259 | 1151 | 160 | 7 | 8.3 | 59.6 | <0.0001 |
| DHS-FHSRR | 19 | 24 | 1110 | 16 | 68 | 0.8 | 52.4 | <0.0001 |

hESC, human embryonic stem cells; HSGRL, human-specific genomic regulatory loci; HSTFBS, human-specific transcription factor-binding sites; HARs, human accelerated regions; revTADs, rapidly-evolving in humans topologically-associating domains; FHSRR, fixed human-specific regulatory regions; DHS, DNase hypersensitive sites; Expected number of genomic features was estimated based on the ratio of the number of human rapidly-evolving TAD to the total number of TAD in hESC (n = 3,127);

Table 2. Genomic features associated with sixty rapidly – evolving in humans topologically - associating domains

| Genomic features | Genome | Rapidly-evolving TADs | Expected | Enrichment | P-value |
|---|---|---|---|---|---|
| Human Accelerated Regions (HARs) | 2,745 | 378 | 53 | 7.4 | < 0.0001 |
| Human-specific TFBS | 3,803 | 1,370 | 73 | 18.8 | < 0.0001 |
| Lamina-associated domains (LADs) | 1,344 | 54 | 26 | 2.1 | 0.0019 |
| Human-specific CTCF-binding sites | 591 | 312 | 11 | 28.4 | < 0.0001 |
| Human-specific NANOG-binding sites | 826 | 192 | 16 | 12 | < 0.0001 |
| Human-specific RNAPII-binding sites | 290 | 181 | 6 | 30.2 | < 0.0001 |
| Human-specific regulatory regions identified in H1-hESC | 1,932 | 109 | 37 | 2.9 | < 0.0001 |
| Human-specific regulatory regions identified in multiple cells | 4,249 | 417 | 82 | 5.1 | < 0.0001 |
| DHS-defined human-specific regulatory regions | 2,118 | 558 | 41 | 13.6 | < 0.0001 |
| Human-specific conservative deletions (CONDELs) | 583 | 29 | 11 | 2.6 | < 0.0001 |
| Human ESC enhancers | 6,823 | 240 | 131 | 1.8 | < 0.0001 |
| Human-specific transcriptional network in the brain | 6,622 | 147 | 127 | 1.2 | 0.3856 |
| Primate-specific CTCF-binding sites | 29,081 | 1,269 | 558 | 2.3 | < 0.0001 |
| H3K27ac peaks with human-specific enrichment in embryonic limb at E33 stage | 780 | 31 | 15 | 2.1 | 0.0238 |
| H3K4me3 peaks with human-specific enrichment in prefrontal cortex (PFC) neurons | 410 | 29 | 8 | 3.6 | < 0.0001 |

hESC, human embryonic stem cells; TFBS, transcription factor-binding site; HARs, human accelerated region; LAD, lamina-associated domain; TAD, topologically-associating domain; RNAPII, RNA polymerase II; PFC, prefrontal cortex; DHS, DNase hypersensitive sites; CONDELs, conservative deletions; E33, embryonic day 33; Expected number of genomic features was estimated based on the ratio of the number of human rapidly-evolving TADs (n = 60) to the total number of TADs in hESC (n = 3,127);

Table 3. Association of hESC super-enhancers with human-specific regulatory loci

| Genomic Features* | TADs | H1 hESC-SE | hsTFBS | HARs |
|---|---|---|---|---|
| Genome | 3,062 | 642 | 3803 | 2745 |
| HARs + SE | 168 | 212 | 0 | 406 |
| Enrichment Factor | 1.0 | 6.0 | NA | 2.8 |
| HSTFBS + SE | 56 | 72 | 146 | 0 |
| Enrichment Factor | 1.0 | 6.1 | 2.1 | NA |
| HARs + HSTFBS + SE | 55 | 85 | 154 | 158 |
| Enrichment Factor | 1.0 | 7.4 | 2.3 | 3.3 |
| SE + HSGRL | 279 | 369 | 300 | 564 |

HARs, Human Accelerated Regions; HSTFBS, human-specific Transcription Factor-Binding Sites; SE, Super Enhancers; HSGRL, human-specific genomic regulatory loci; H1 hESC, H1 human Embryonic Stem Cells; NA, not applicable; * - genomic coordinates of the hg19 human genome reference database;

Table 4. Size distribution analysis of super-enhancers in human and mouse ESC

| Category | hESC super-enhancers | | mESC super-enhancers | | |
|---|---|---|---|---|---|
| Size, bp | Number | Percent | Number | Percent | Enrichment in hESC |
| > 30,000 | 29 | 4.2 | 4 | 1.7 | 7.25 |
| >20,000 | 102 | 14.9 | 28 | 12.1 | 3.64 |
| >10,000 | 331 | 48.4 | 108 | 46.8 | 3.06 |
| >5,000 | 554 | 81.0 | 154 | 66.7 | 3.60 |
| >2,000 | 683 | 99.9 | 193 | 83.5 | 3.54 |
| 1,000-2,000 | 0 | 0.0 | 21 | 9.1 | 0.00 |
| <1,000 | 1 | 0.1 | 17 | 7.4 | 0.06 |
| Total | 684 | 100.0 | 231 | 100.0 | 2.96 |

Table 5. Size distribution analysis of topologically-associating domains in genomes of human and mouse ESC

| Category | hESC TADs | | mESC TADs | | |
|---|---|---|---|---|---|
| Size, Kbp | Number | Percent | Number | Percent | Enrichment/Depletion in hESC |
| > 3,000 | 23 | 0.7 | 55 | 2.5 | 0.42 |
| >2,000 | 160 | 5.1 | 255 | 11.6 | 0.63 |
| >1,000 | 955 | 30.5 | 982 | 44.6 | 0.97 |
| >500 | 2154 | 68.9 | 1745 | 79.3 | 1.23 |
| >200 | 3054 | 97.7 | 2166 | 98.5 | 1.41 |
| >100 | 3125 | 99.9 | 2200 | 100.0 | 1.42 |
| <100 | 2 | 0.1 | 0 | 0.0 | NA |
| Total | 3127 | 100.0 | 2200 | 100.0 | 1.42 |

Table 6. Estimates of creation time periods during evolution of super-enhancers and enhancers in the hESC genome

1.1. Estimates of super-enhancers (SE) creation time in the human ESC genome

| Number of SE in hESC genome | Shared with mESC | All new SE in hESC* | New SE on conserved sequences* | Primate-specific SE sequences* | Human-specific SE sequences** |
|---|---|---|---|---|---|
| 684 | 25 (3.7%) | 659 | 573 | 83 | 3 |
| Number of new SEs per 100,000 years | | 0.88 | 0.76 | 0.11 | 0.81 |
| Years for creation of one super-enhancer | | 113,636 | 131,579 | 909,091 | 123,457 |

1.2. Estimates of enhancers' creation time in the human ESC genome

| Number of enhancers in hESC genome | Shared with mESC | All new enhancers* | New enhancers on conserved sequences* | Primate-specific enhancers sequences* | Human-specific enhancers sequences** |
|---|---|---|---|---|---|
| 7006 | 682 (9.7%) | 6,324 | 4,714 | 1,420 | 190 |
| New enhancers per 100,000 years | | 8.4 | 6.3 | 1.9 | 51 |
| Years for creation of one enhancer | | 11,905 | 15,873 | 52,632 | 1,961 |

* Calculated based on estimates of Humans & Chimpanzees split 13 million years ago and 88 million years from Euarchonta & Glires (Gliriformes) split resulting in the estimated evolutionary timeline of ~75 million years;
** Calculated based on estimates of Modern Humans & Neanderthals split 370,000 years ago;

   Estimates for human-specific SE and enhancers were calculated based on the assumption that all human-specific sequences emerged after Modern Humans & Neanderthals split 370,000 years ago;
   Estimates of creation time of primate-specific SE and enhancers were based on the assumption that all primate-specific sequences emerged before Modern Humans/Chimpanzees split 13 million years ago.

**Figure 1.**

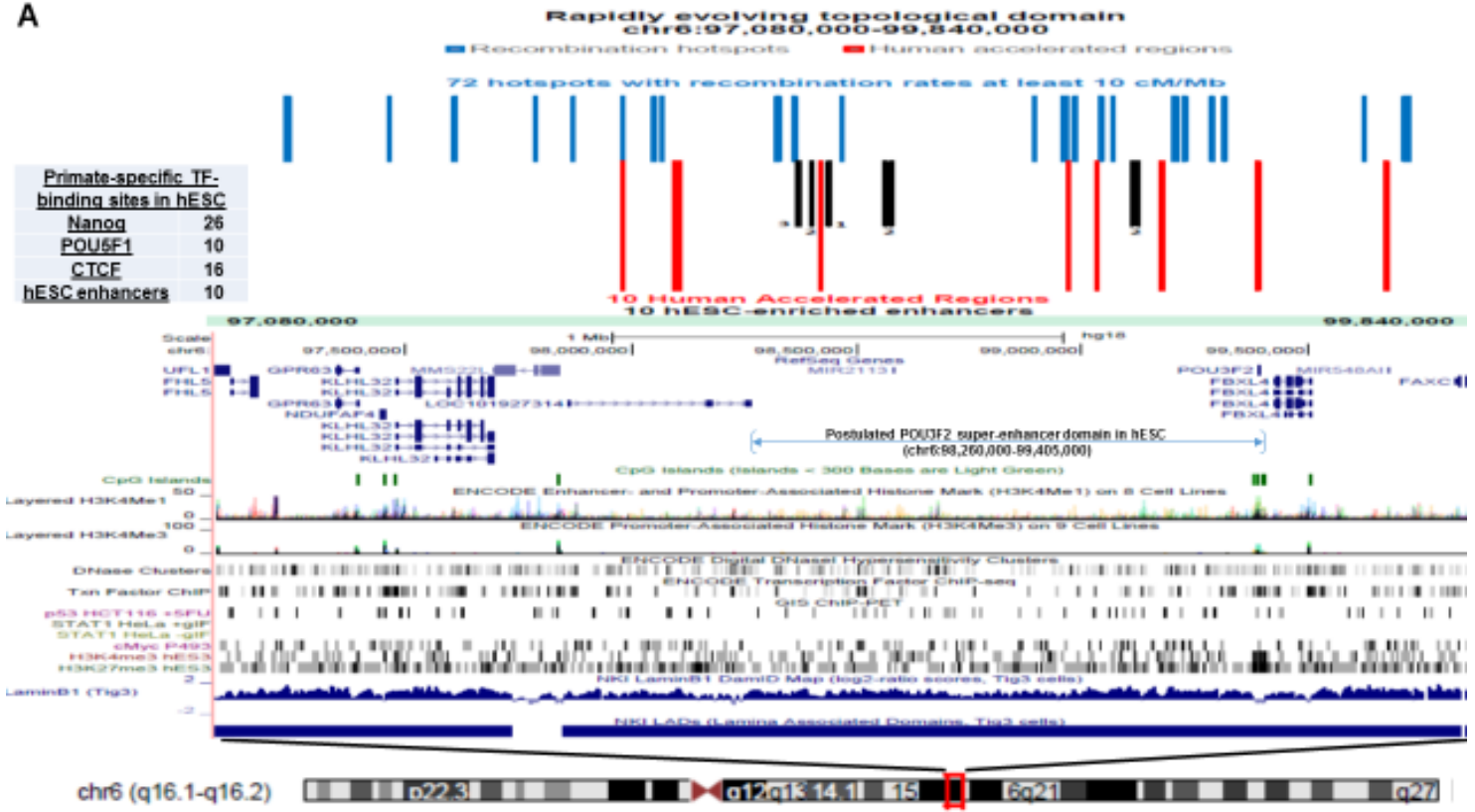
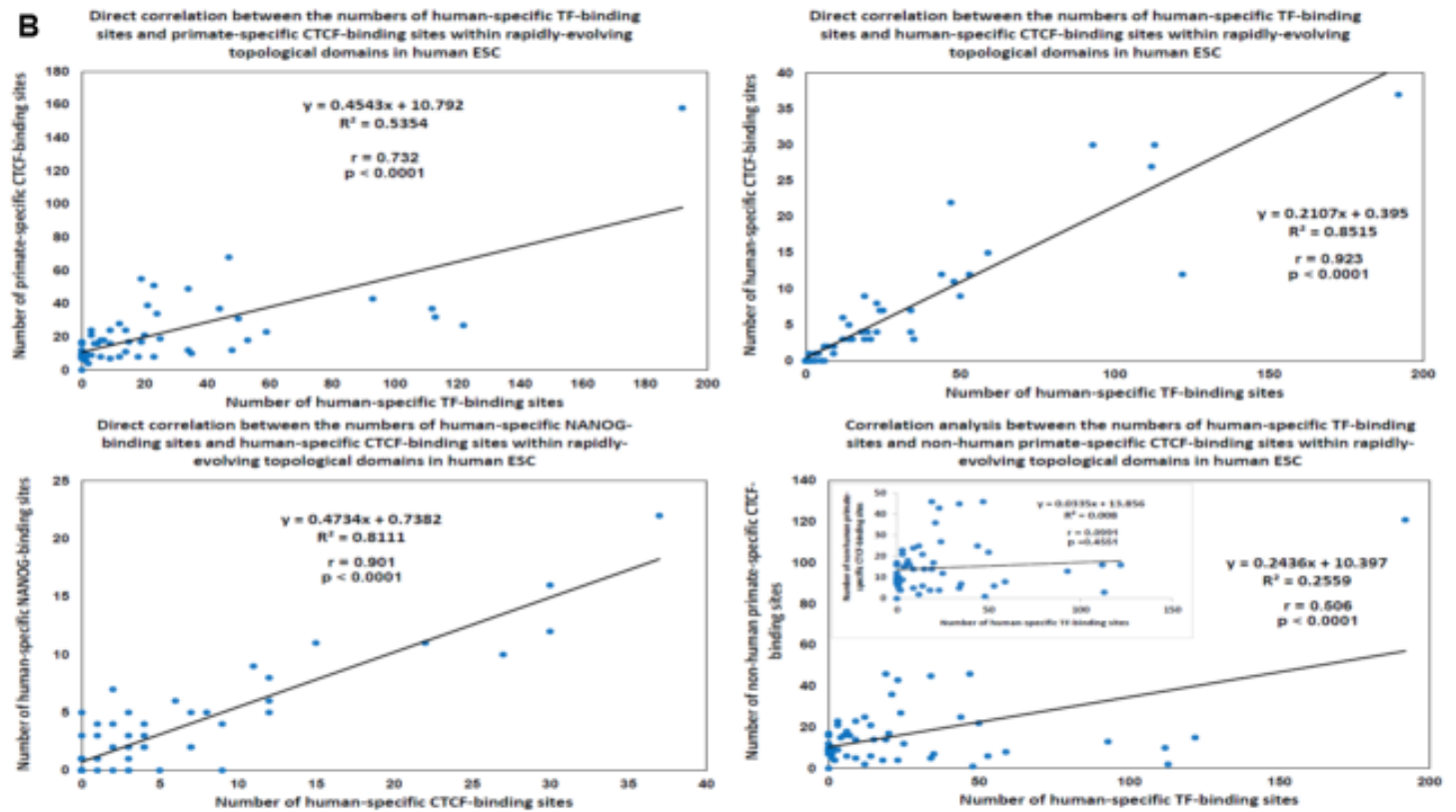

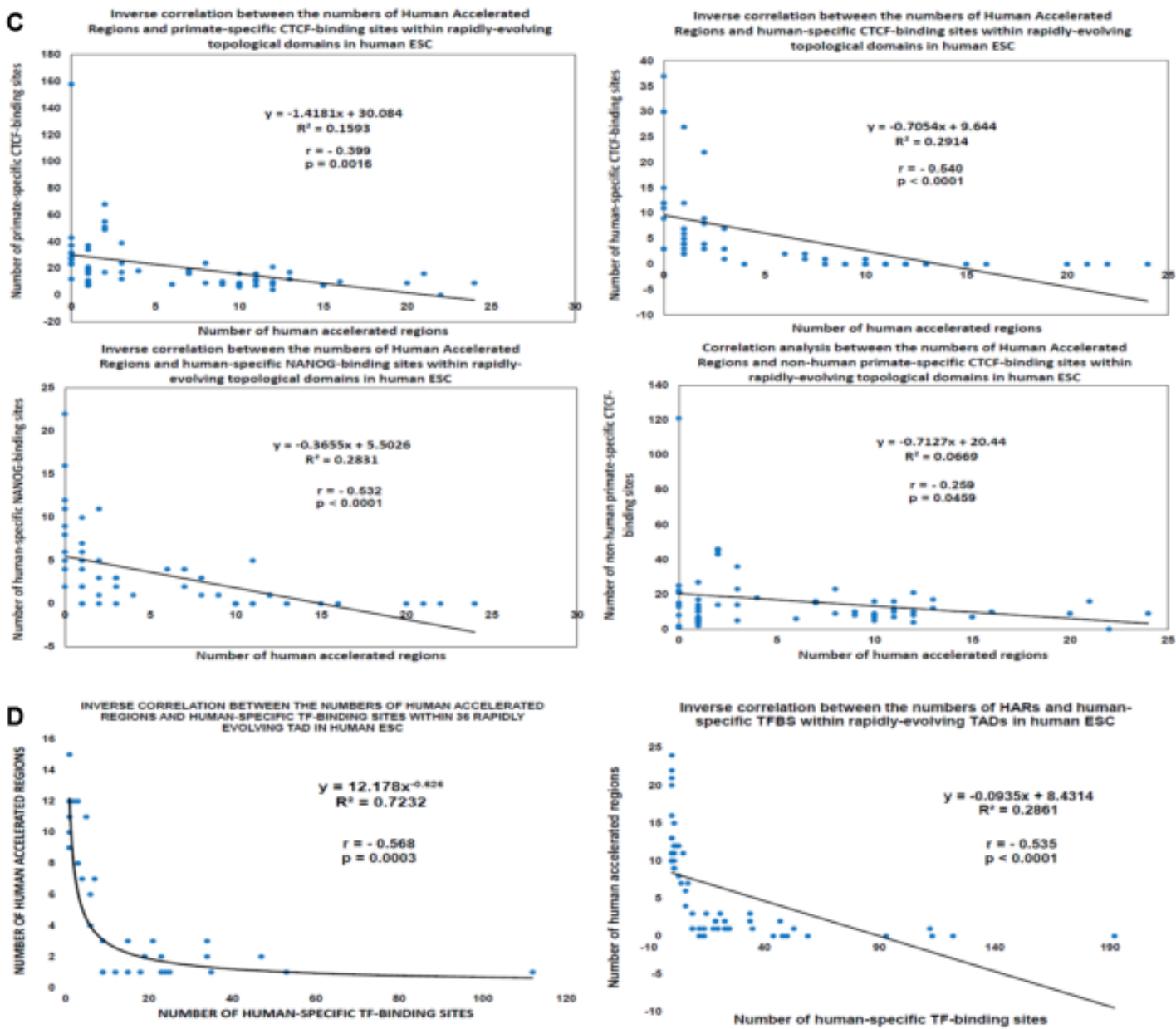

**Figure 2.**

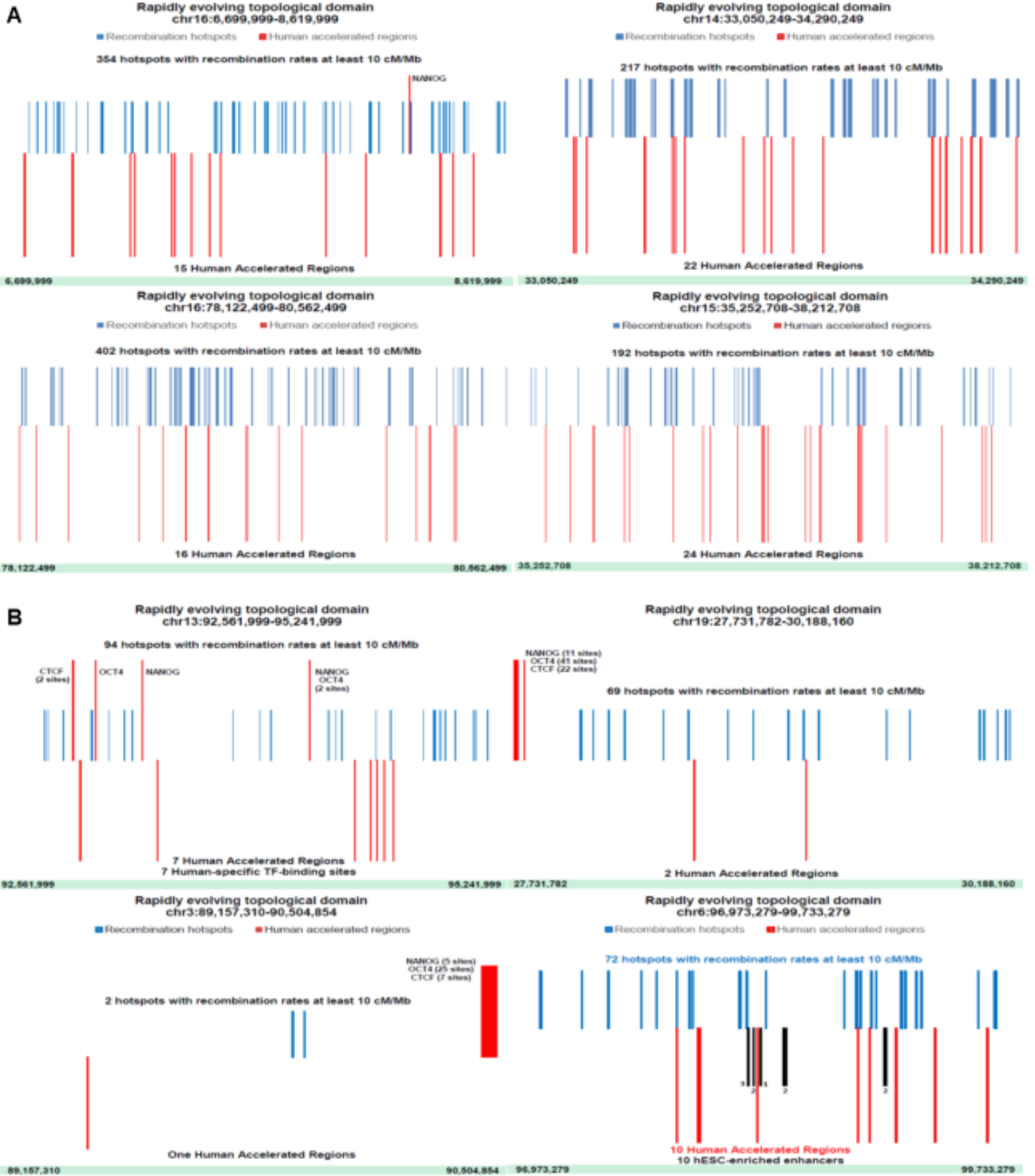

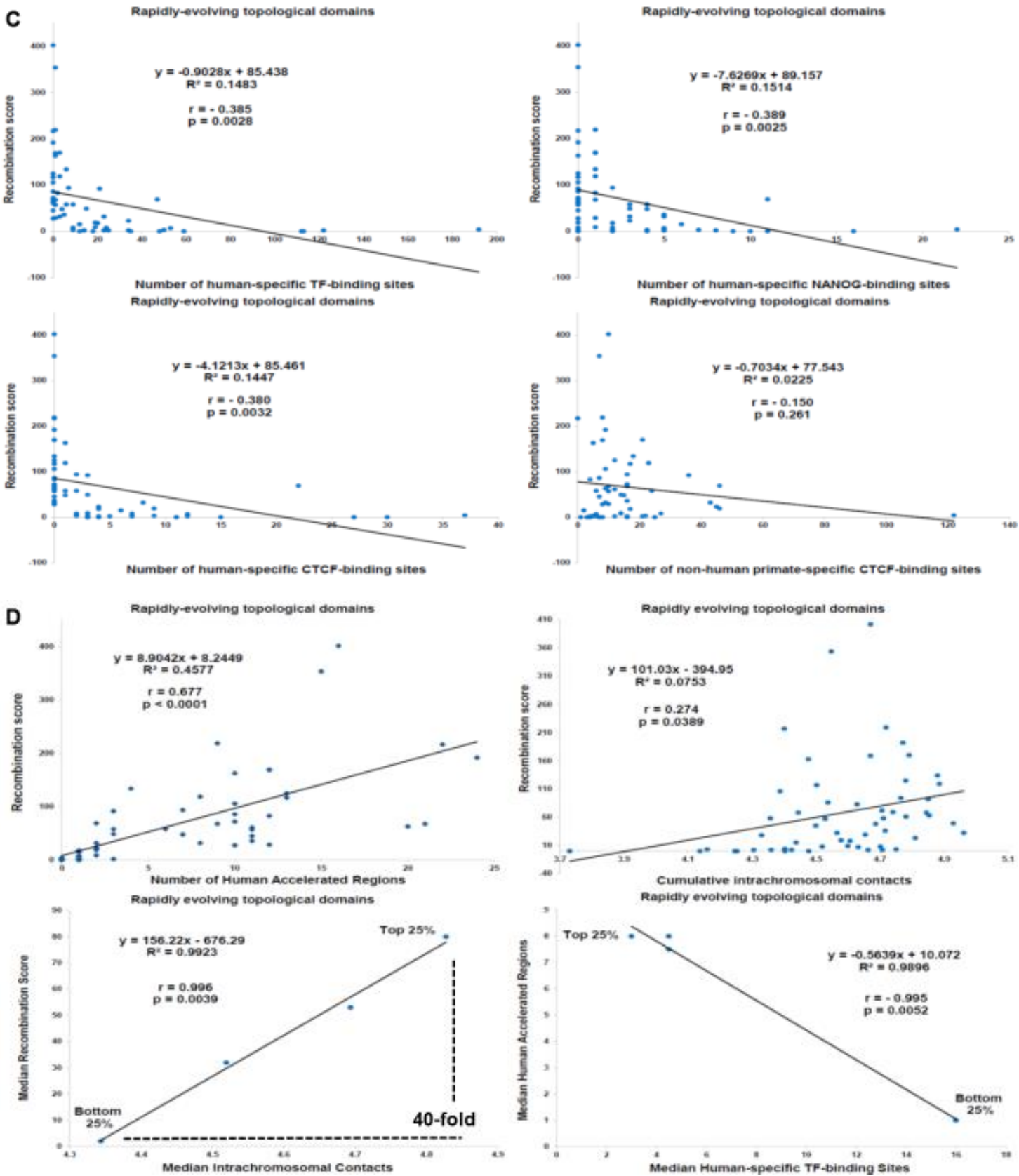

**Figure 3.**

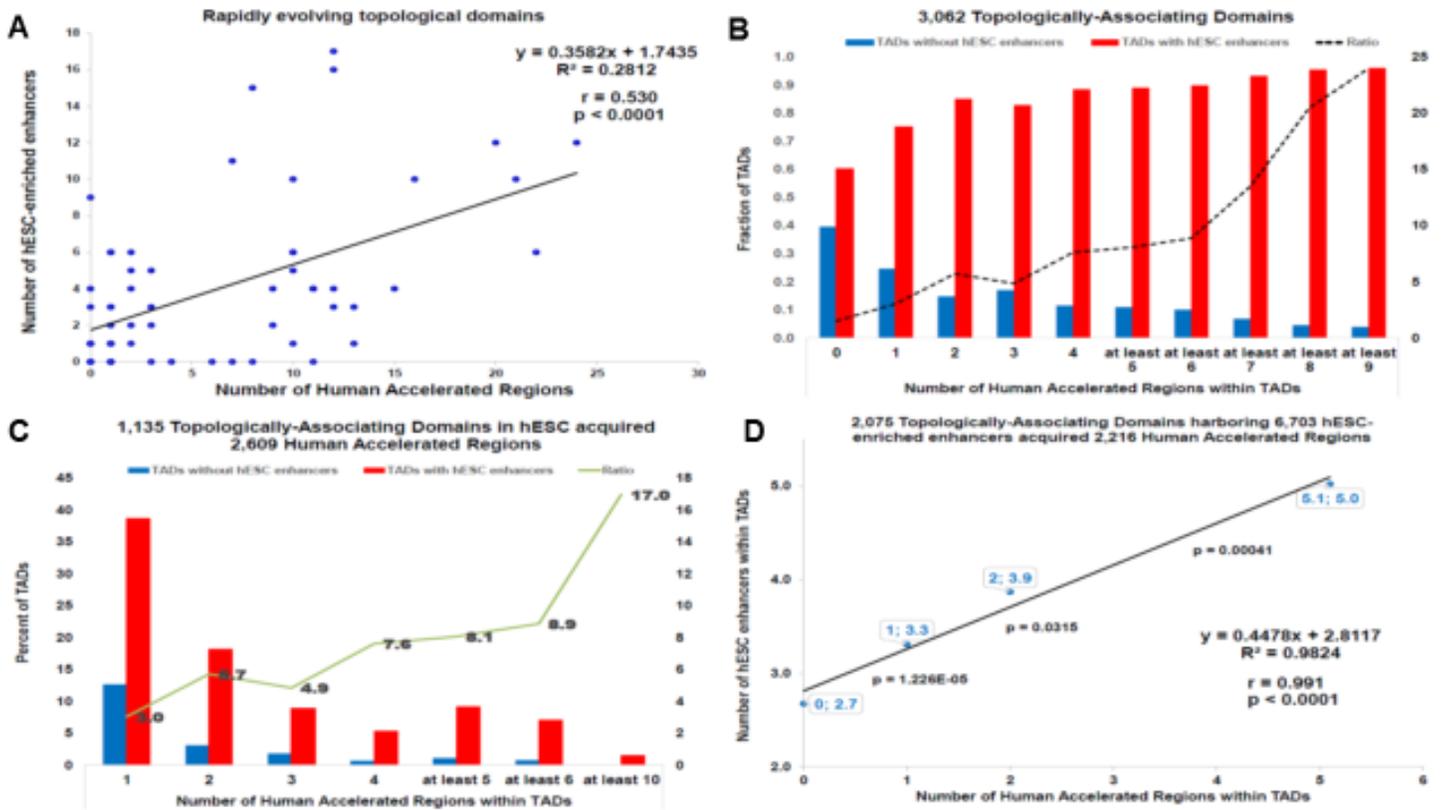

**Figure 4.**

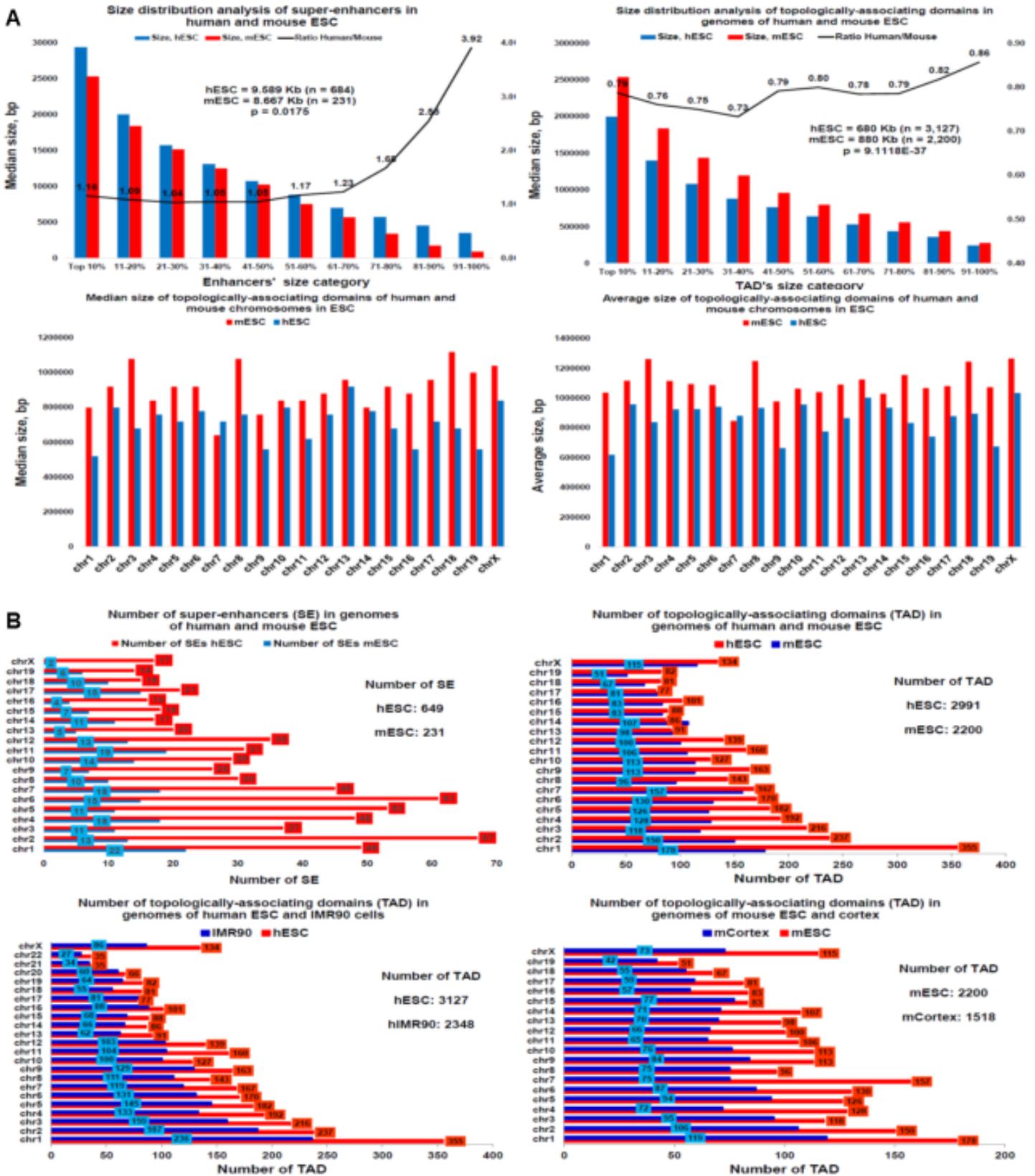

**Figure 5.**

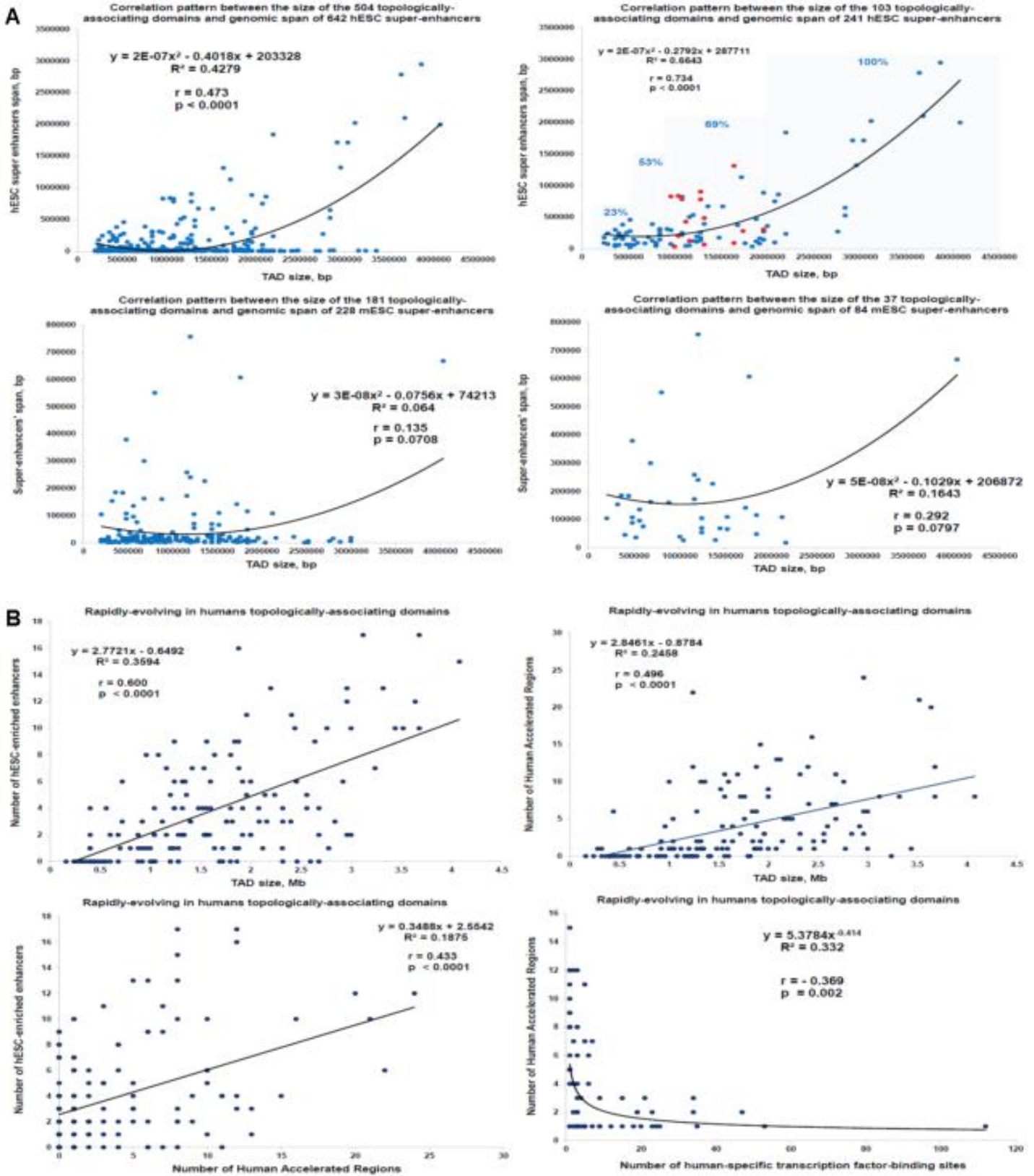

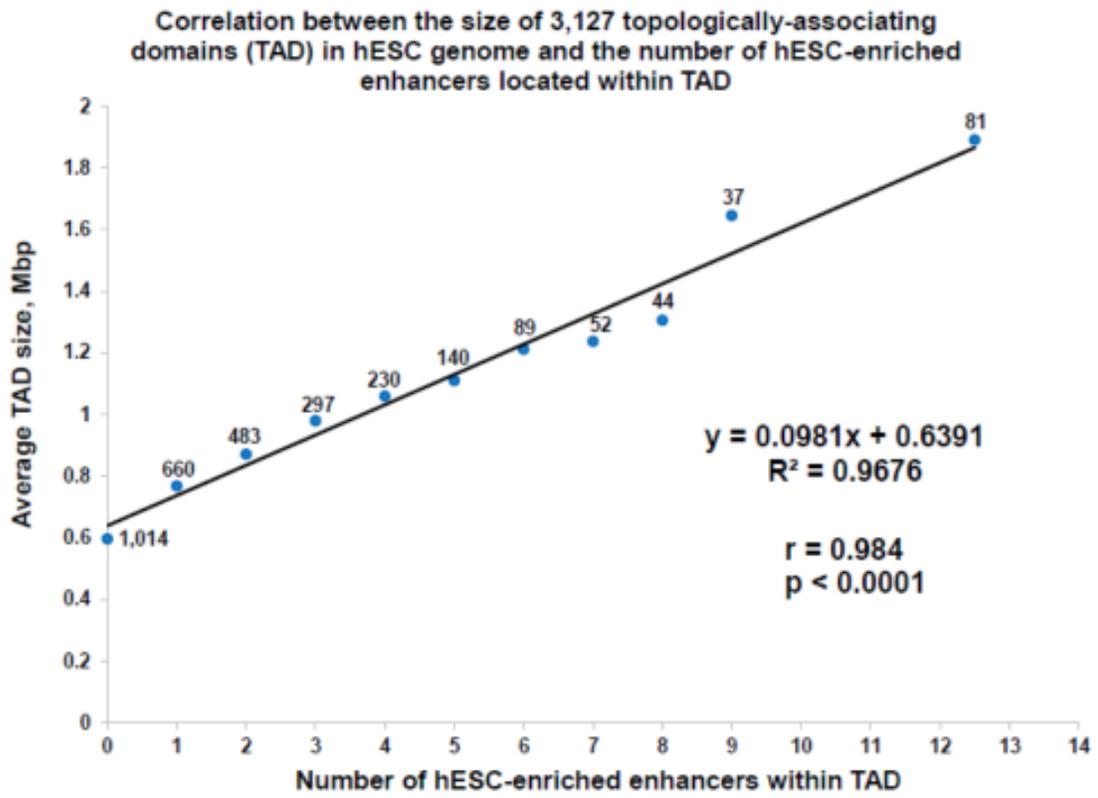

**Figure 6.**

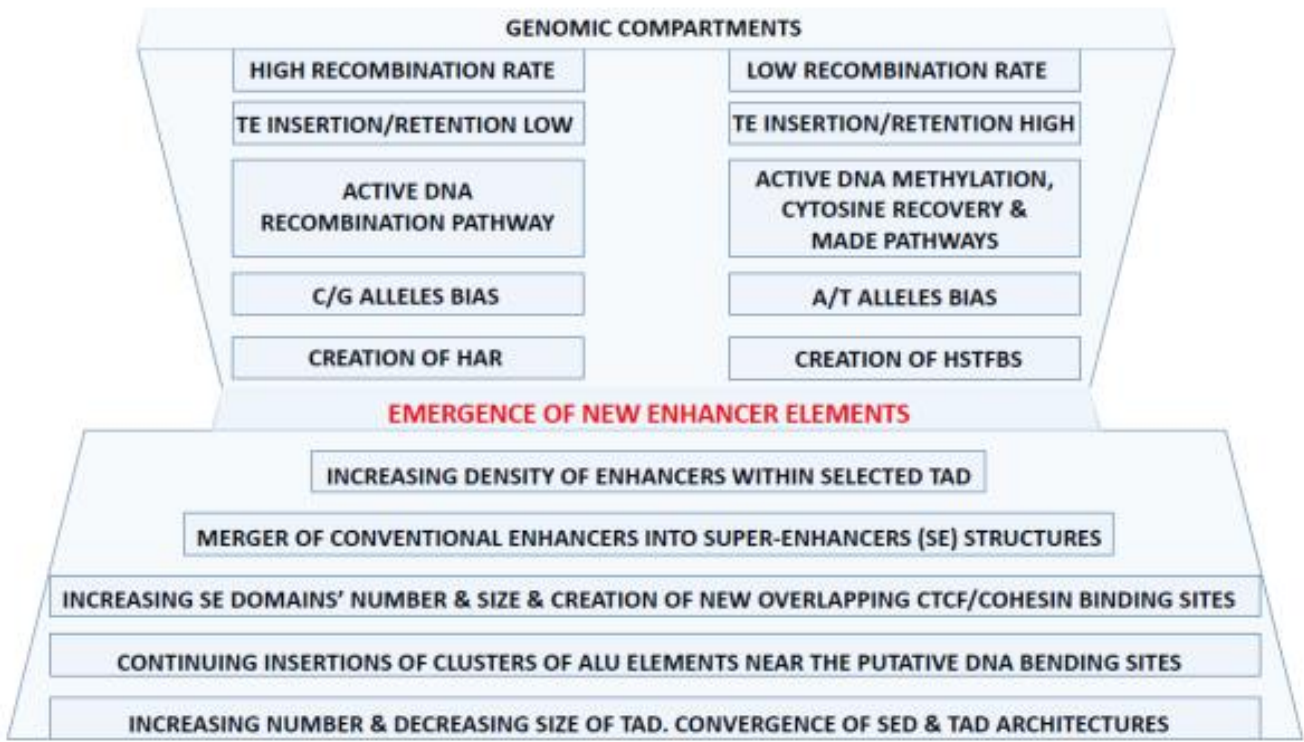

**Figure 7.**

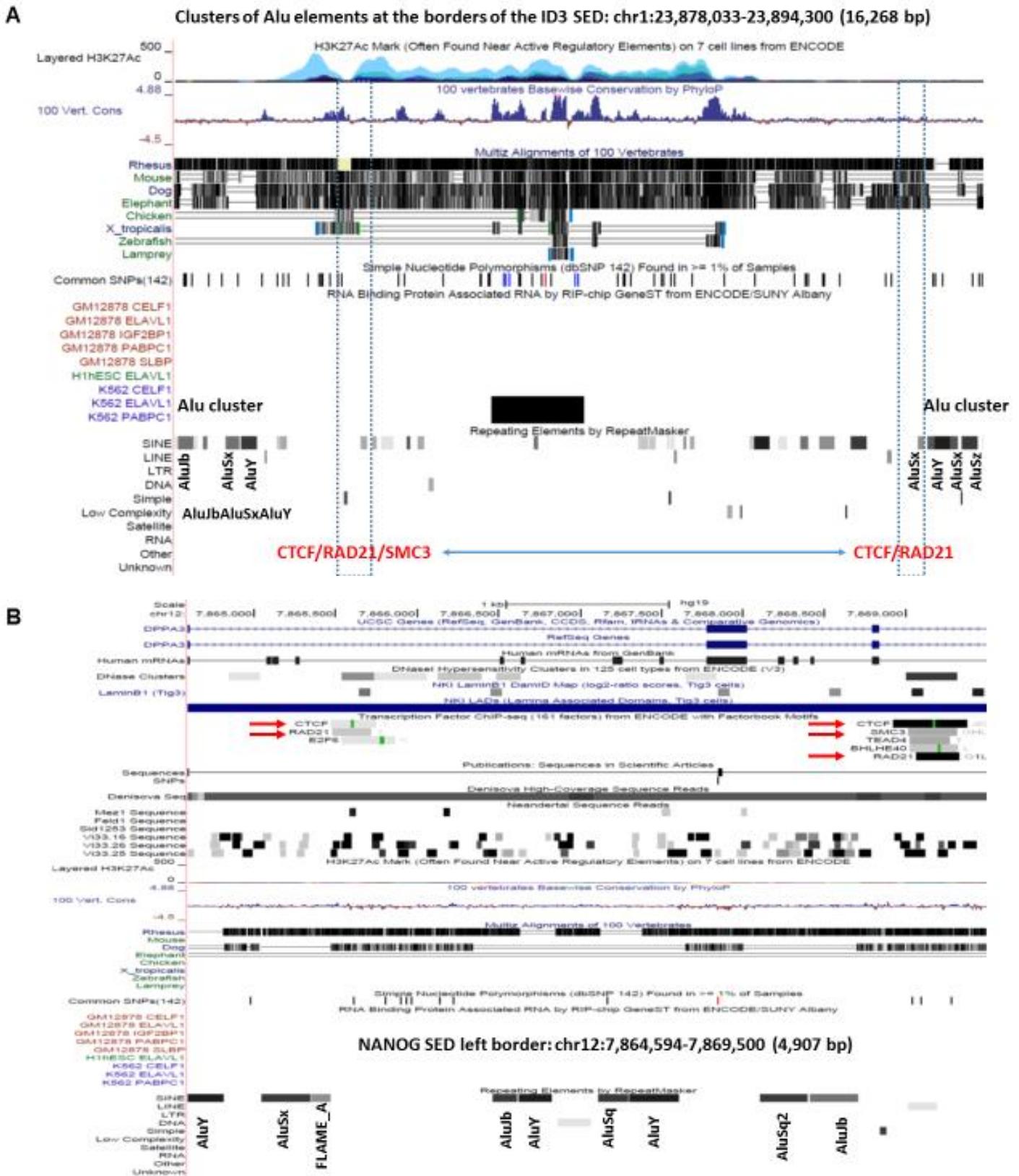

C

NANOG SED right border: chr12:8,012,400-8,017,400 (5,001 bp)

**Figure 8.**

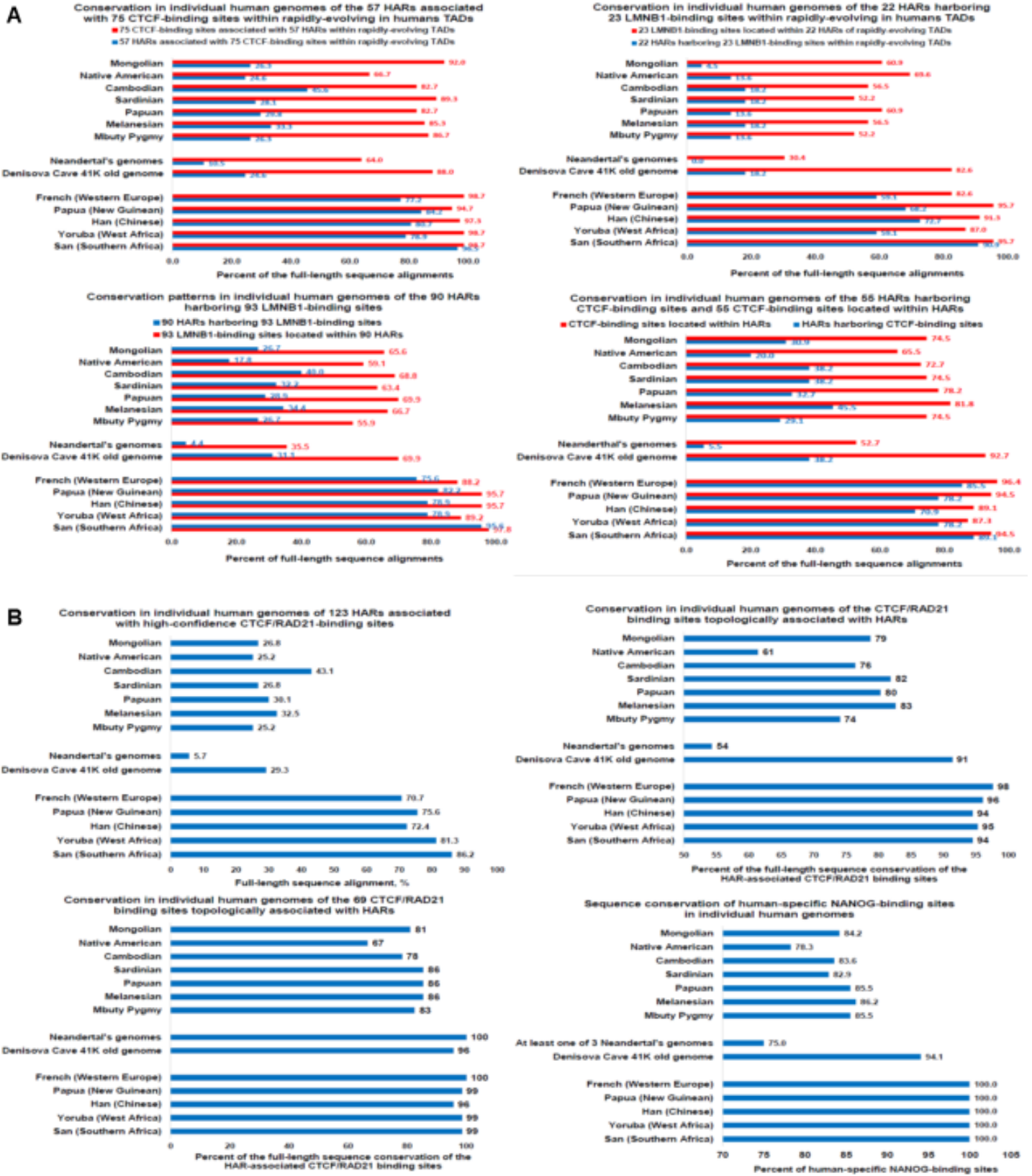

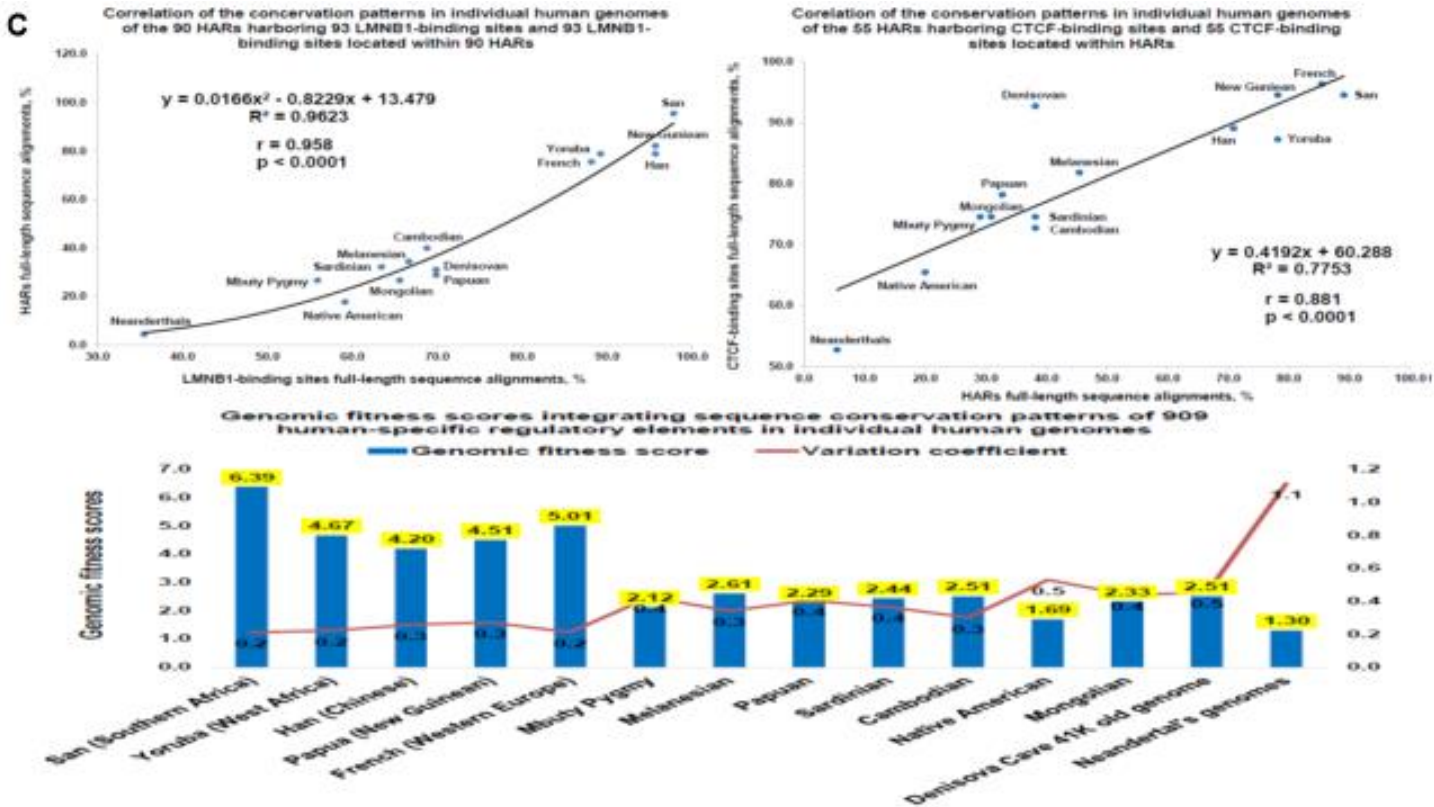

**Figure 9.**

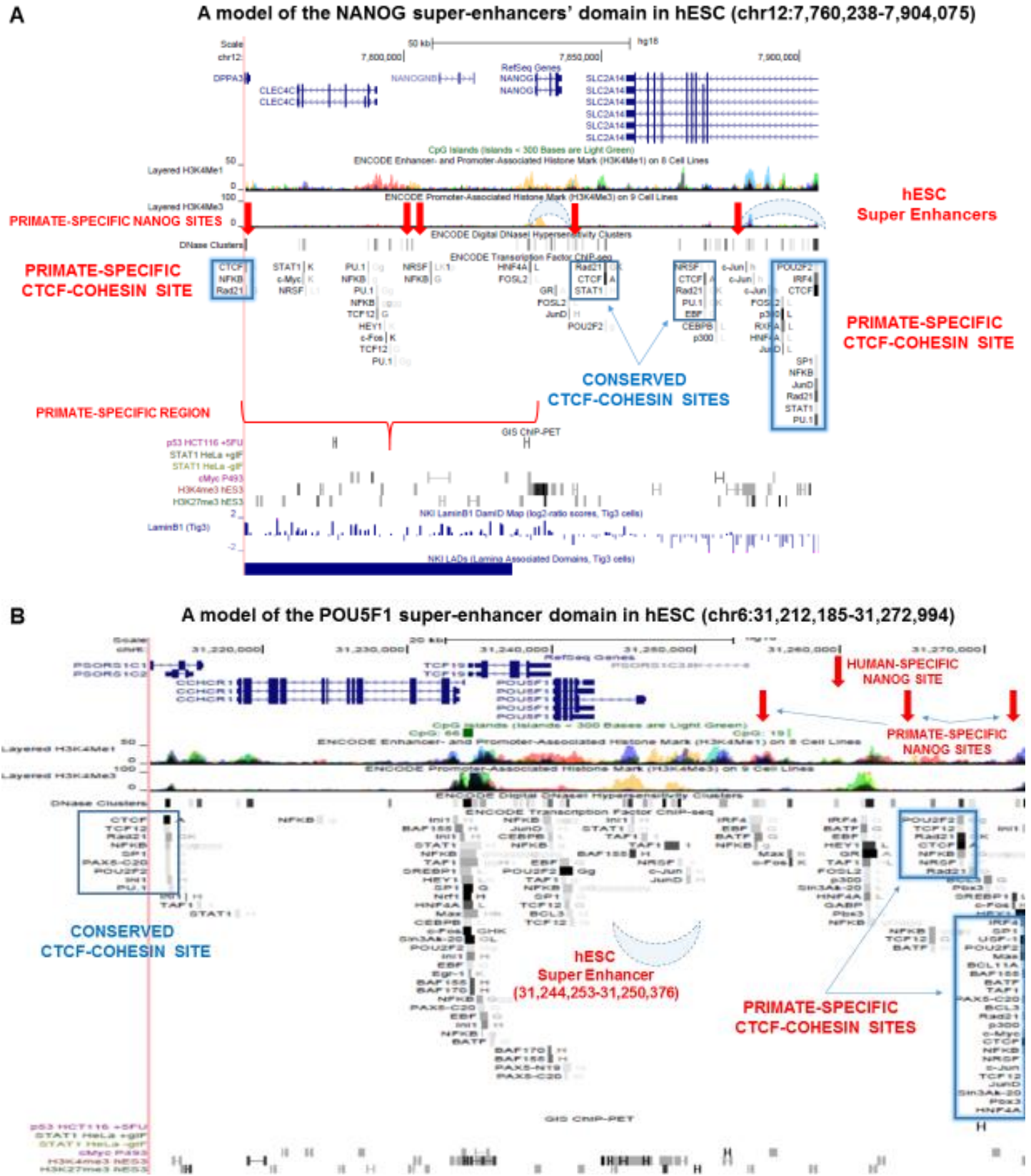

C

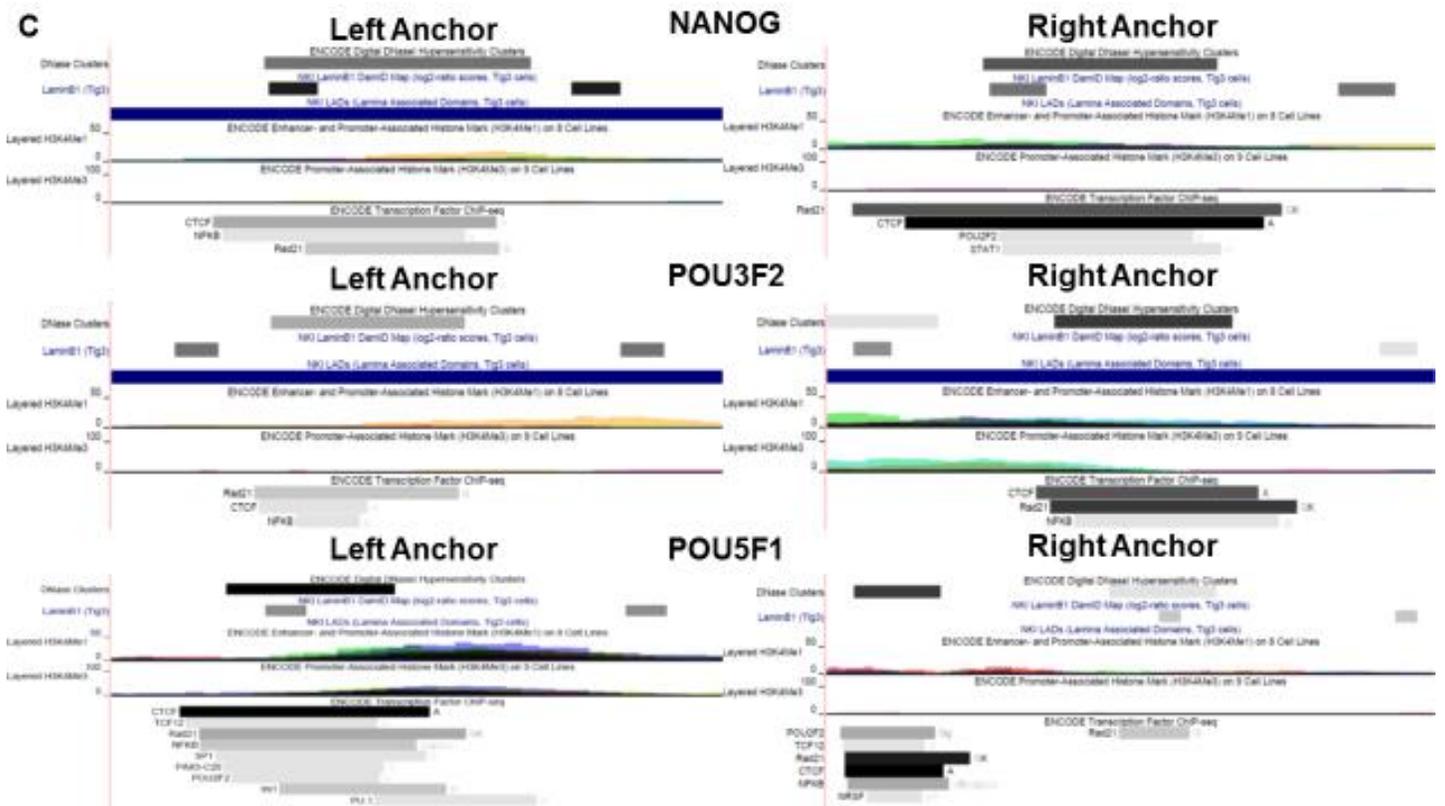

**Figure 10.**

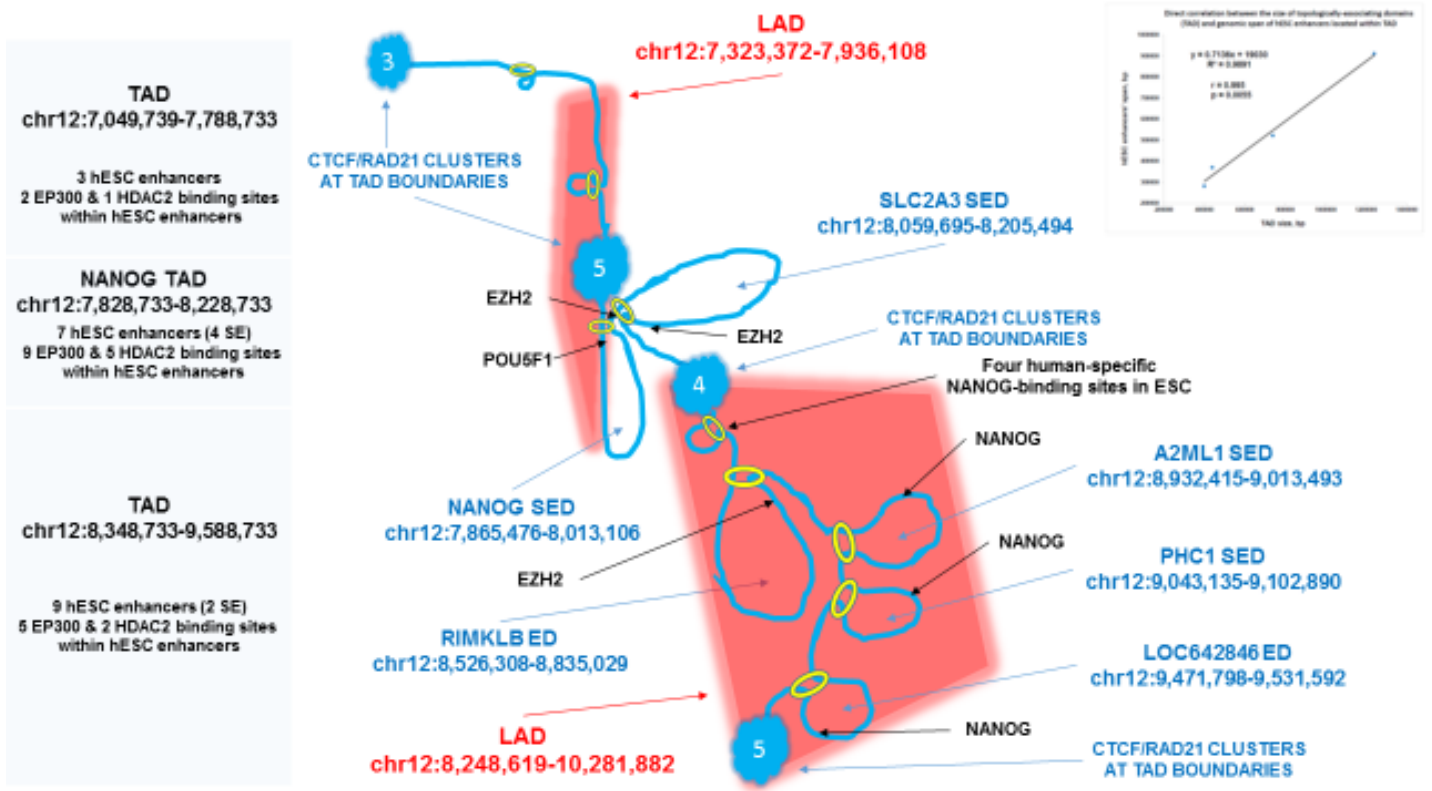

**Figure 11.**

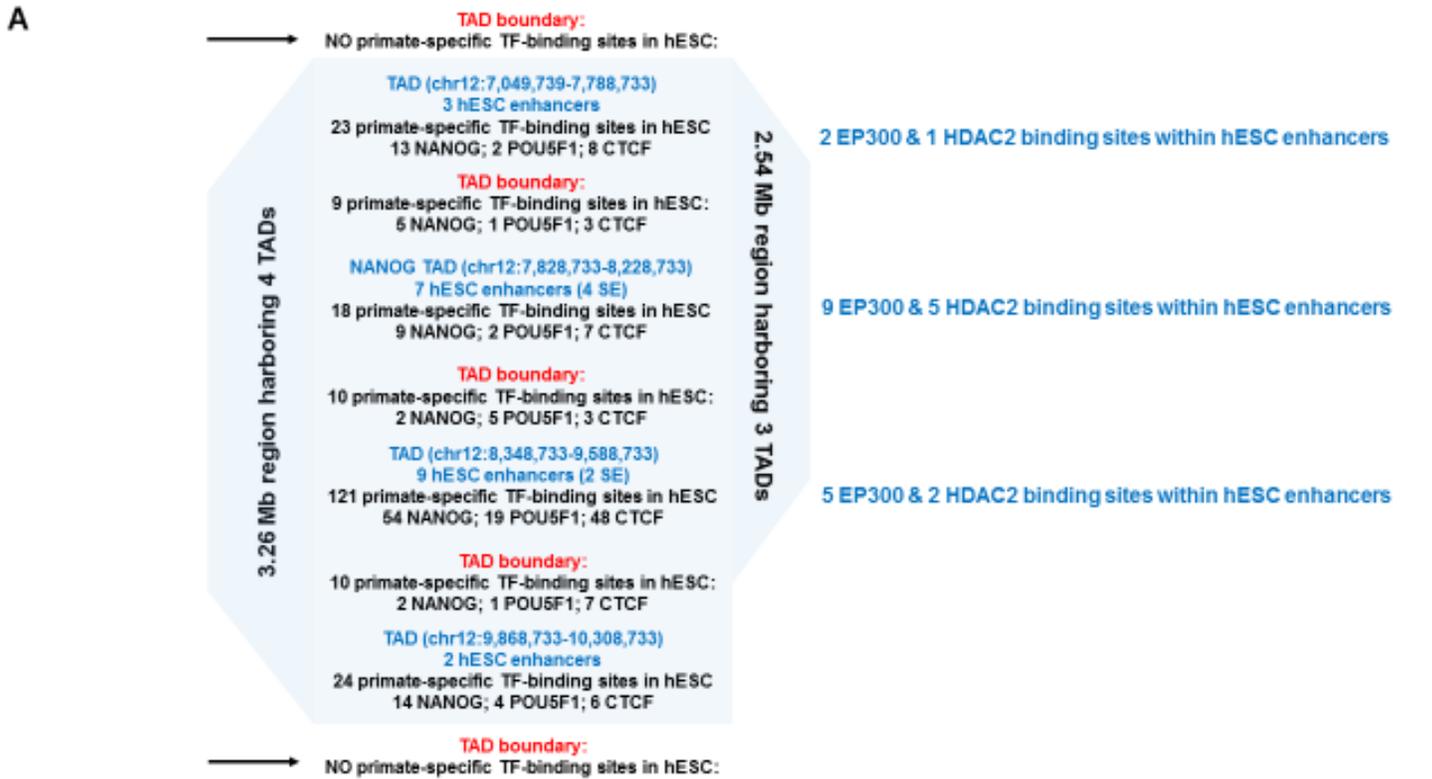

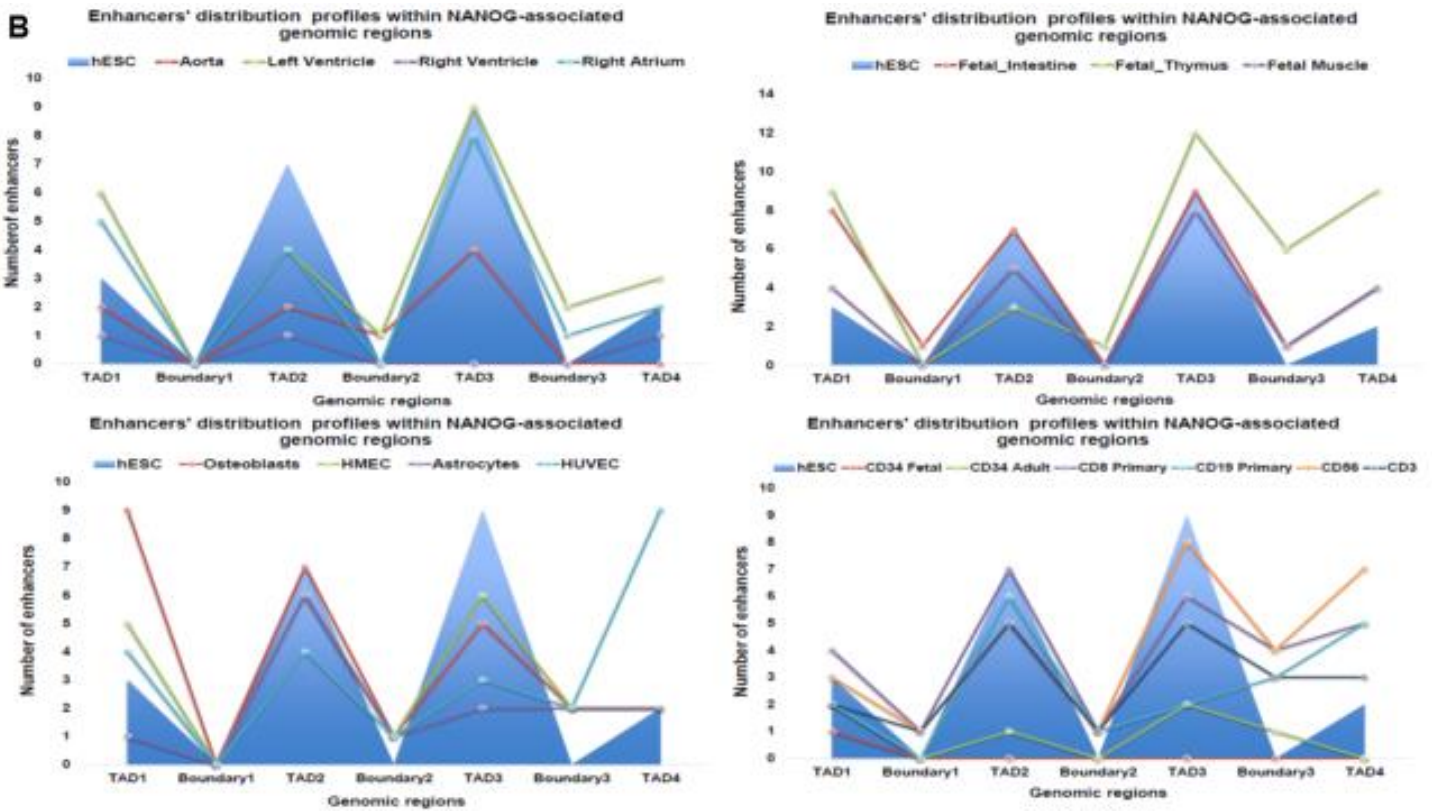

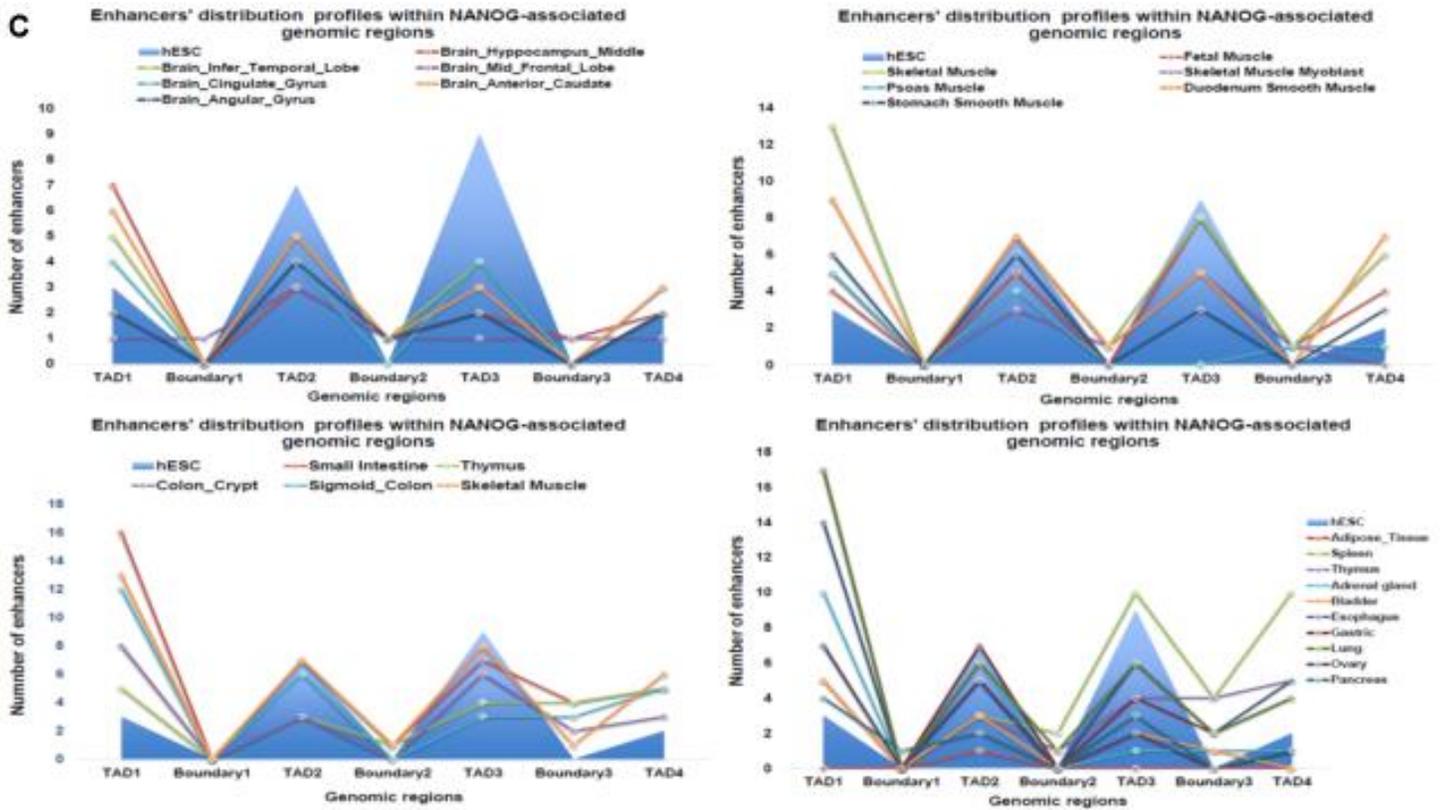
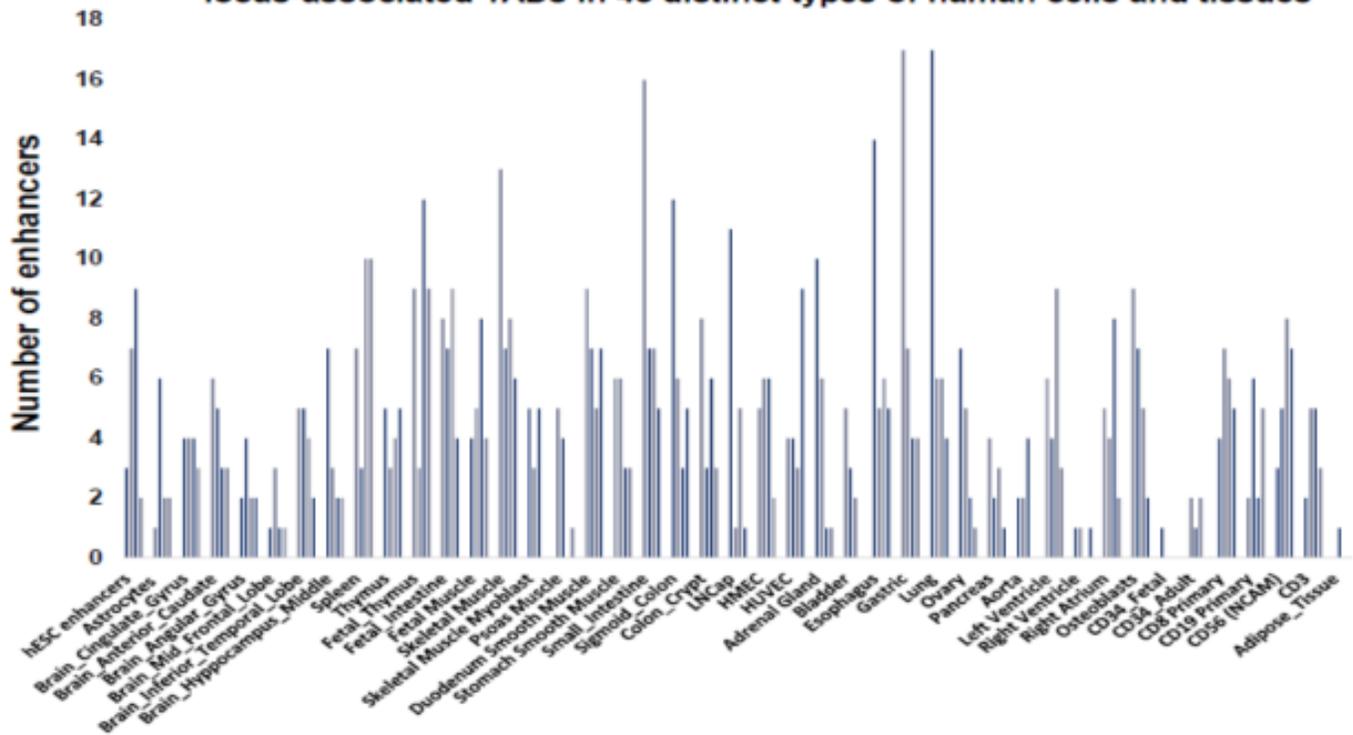

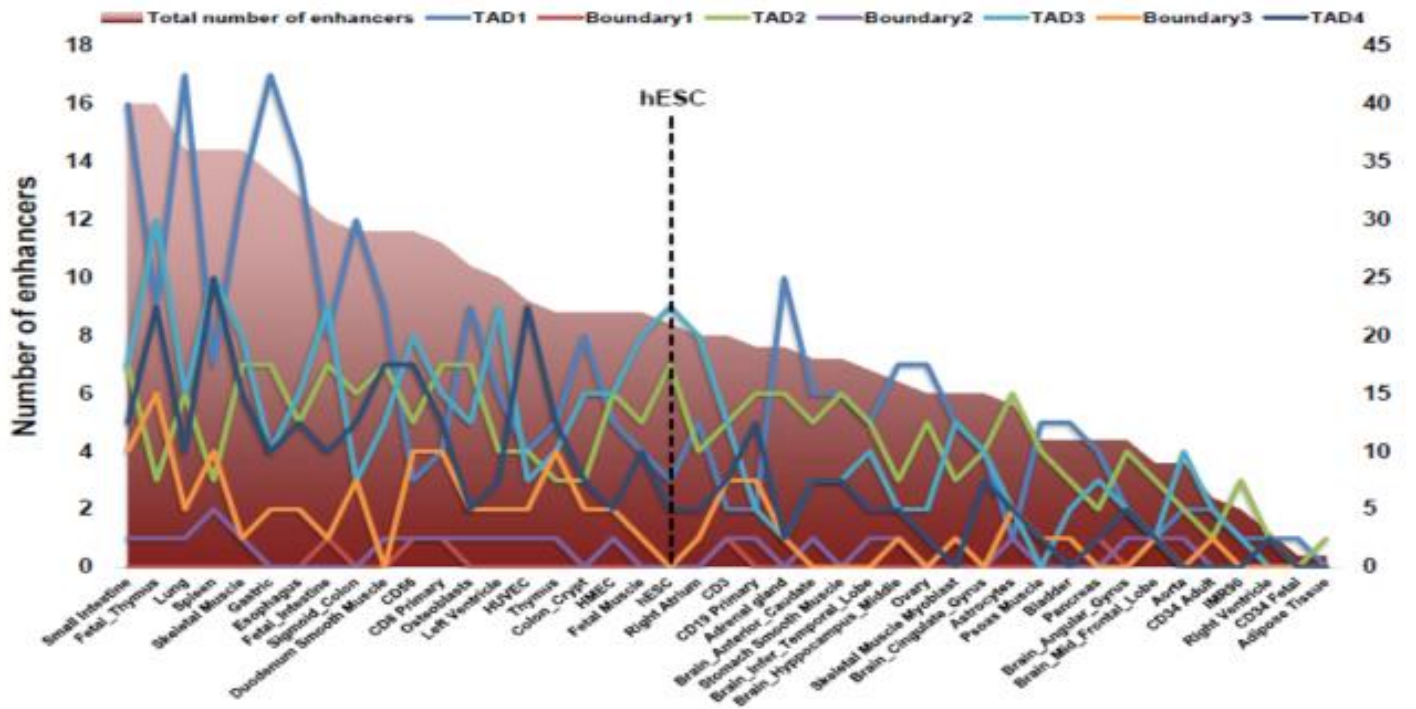

**Figure 12.**

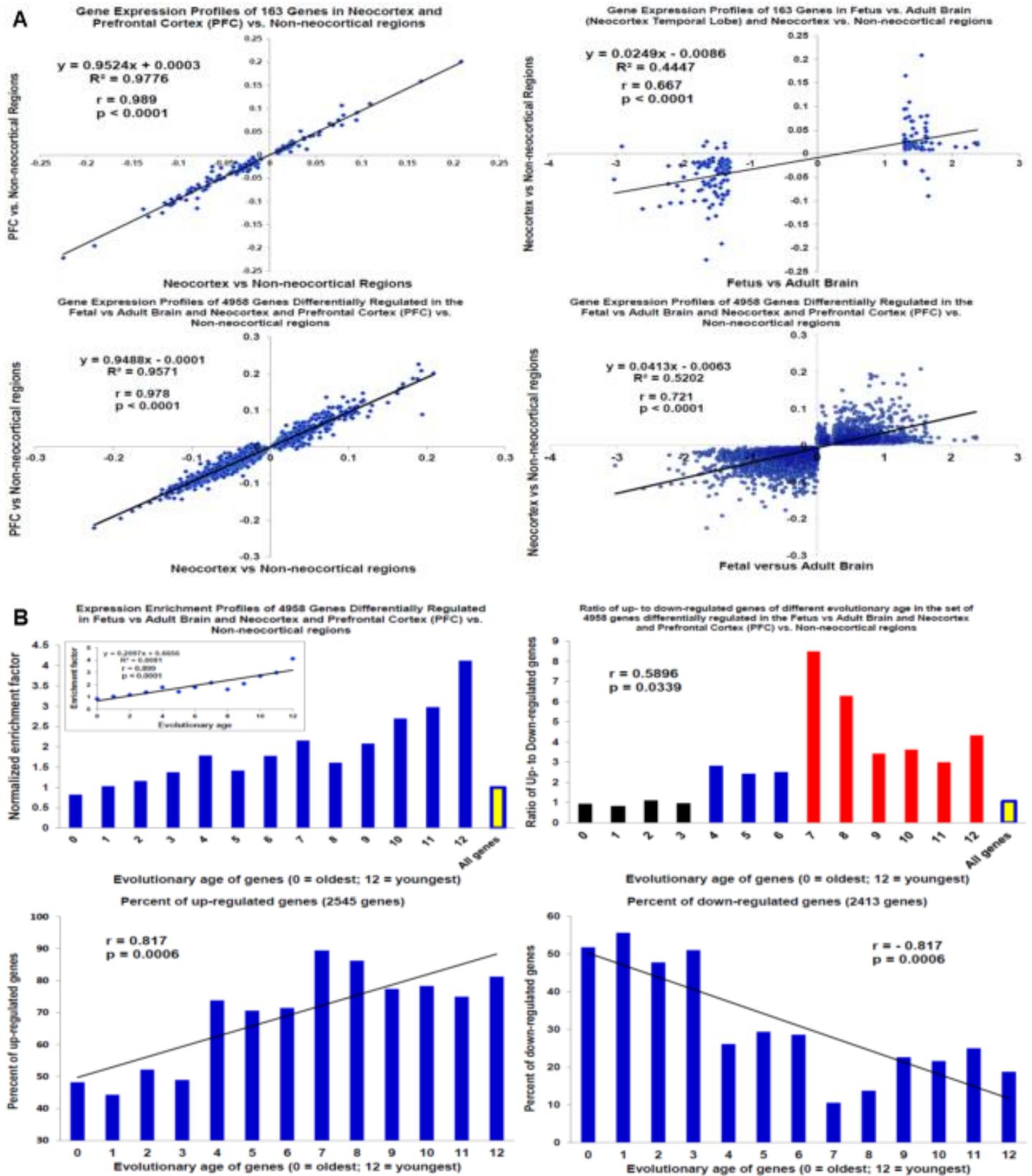

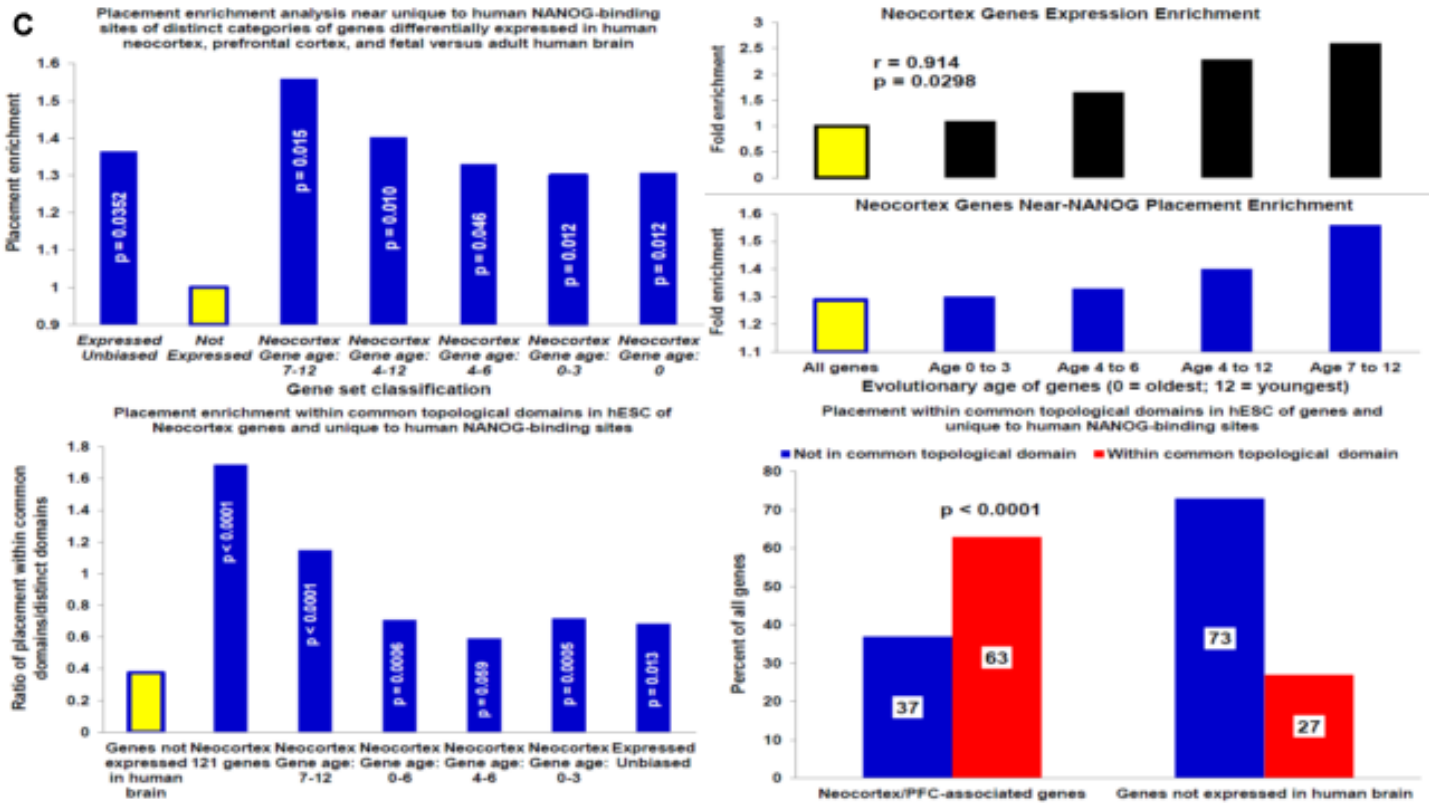

**Figure 13.**

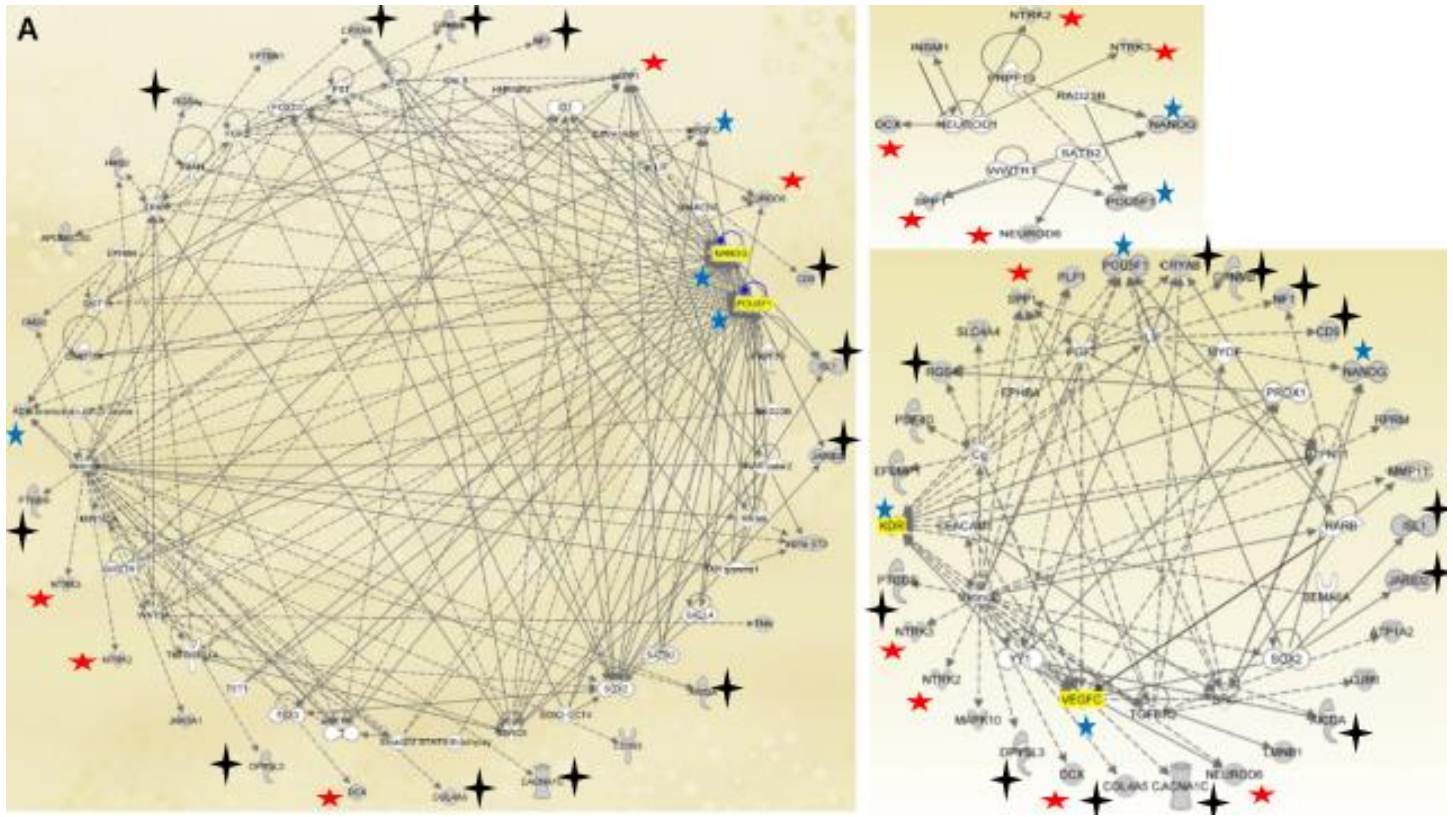

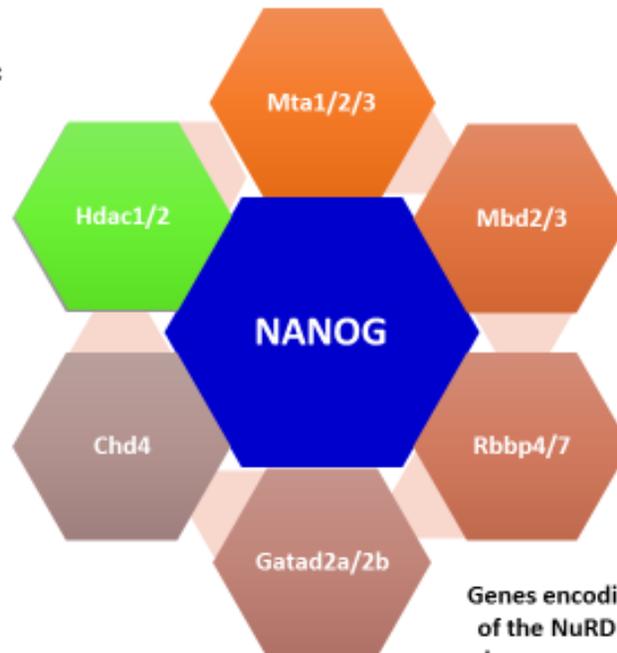